\documentclass[lettersize,journal]{IEEEtran}
\ifCLASSOPTIONcompsoc
  \usepackage[nocompress]{cite}
\else
  \usepackage{cite}
\fi
\ifCLASSINFOpdf
\usepackage[pdftex]{graphicx}

\else

\fi

\usepackage{cite}
\usepackage{amsmath,amssymb,amsfonts}
\usepackage{algorithmic}
\usepackage{algorithm}
\usepackage{tcolorbox}
\usepackage{multirow}
\usepackage{tabularx}   % for X column
\usepackage{graphicx}
\usepackage{textcomp}
\usepackage[normalem]{ulem}
\usepackage[dvipsnames]{xcolor}
\usepackage{url}
\usepackage{subfig}
\usepackage{booktabs}   % for toprule/midrule/bottomrule
\usepackage{xcolor}
\usepackage{soul}

\usepackage{color}
 
\usepackage{amsmath}

\usepackage{ulem}
\usepackage{float}
\usepackage{makecell}

\usepackage{multirow}
\usepackage{subfloat}
\usepackage{subfig}
\usepackage{soul}

\usepackage{tikz}  % 为 \pgfmath 提供支持

\usepackage{xcolor}
\usepackage{colortbl}

\usepackage[english]{babel}
\newtheorem{theorem}{Proposition}

\definecolor{bestgreen}{RGB}{190,255,190}   % 最好（增强）
\definecolor{goodgreen}{RGB}{215,255,215}   % 次好（增强）
\definecolor{midwhite}{RGB}{235,255,235}    % 极浅绿（原 midwhite 替换）
\definecolor{badred}{RGB}{255,220,220}      % 次差（增强）
\definecolor{worstred}{RGB}{255,195,195}    % 最差（增强）

\newcommand{\cBest}[1]{\cellcolor{bestgreen}{#1}}
\newcommand{\cGood}[1]{\cellcolor{goodgreen}{#1}}
\newcommand{\cMid}[1]{\cellcolor{midwhite}{#1}}
\newcommand{\cBad}[1]{\cellcolor{badred}{#1}}
\newcommand{\cWorst}[1]{\cellcolor{worstred}{#1}}

\definecolor{G1}{RGB}{190,255,190}
\definecolor{G2}{RGB}{205,255,205}
\definecolor{G3}{RGB}{220,255,220}
\definecolor{G4}{RGB}{235,255,235}
\definecolor{G5}{RGB}{245,255,245}
\definecolor{G6}{RGB}{252,255,252}

\definecolor{R1}{RGB}{255,195,195}
\definecolor{R2}{RGB}{255,210,210}
\definecolor{R3}{RGB}{255,225,225}
\definecolor{R4}{RGB}{255,238,238}
\definecolor{R5}{RGB}{255,246,246}
\definecolor{R6}{RGB}{255,252,252}

\newcommand{\cGOne}[1]{\cellcolor{G1}{#1}}
\newcommand{\cGTwo}[1]{\cellcolor{G2}{#1}}
\newcommand{\cGThree}[1]{\cellcolor{G3}{#1}}
\newcommand{\cGFour}[1]{\cellcolor{G4}{#1}}
\newcommand{\cGFive}[1]{\cellcolor{G5}{#1}}

\newcommand{\cROne}[1]{\cellcolor{R1}{#1}}
\newcommand{\cRTwo}[1]{\cellcolor{R2}{#1}}
\newcommand{\cRThree}[1]{\cellcolor{R3}{#1}}
\newcommand{\cRFour}[1]{\cellcolor{R4}{#1}}

\begin{document}

% \title{CARE for Life: A Wearable Sensor Enhanced Cross-Modal Approach for Robust Glucose Prediction}
% \title{An Interpretable Framework for Predicting and Generating converted conditions of Conditional Potential Users}

\title{Bringing GRACE to Recommendation: Fine-Tuning for Sustainable and Accurate Personalization}

\author{
\IEEEauthorblockN{
Yibowen Zhao \IEEEmembership{Graduate Student Member, IEEE},
Yinan Zhang,
Ning Liu, \\
Lizhen Cui \IEEEmembership{Senior Member, IEEE},
Chunyan Miao \IEEEmembership{Fellow, IEEE}
}

\thanks{Yibowen Zhao, Ning Liu and Lizhen Cui are with the Joint SDU-NTU Centre for Artificial Intelligence Research (C-FAIR) and School of Software, Shandong University, Jinan, China (e-mail: ybw.zhao@mail.sdu.edu.cn; liun21cs@sdu.edu.cn; clz@sdu.edu.cn).}

\thanks{Yinan Zhang and Chunyan Miao are with the Alibaba-NTU Global e-Sustainability CorpLab (ANGEL), Nanyang Technological University, Singapore (e-mail: yinan.zhang@ntu.edu.sg; ascymiao@ntu.edu.sg).}

\thanks{\IEEEauthorrefmark{1}Corresponding authors.}
}

\markboth{IEEE Transactions on Knowledge and Data Engineering}%
{Shell \MakeLowercase{\textit{et al.}}: Bare Demo of IEEEtran.cls for Computer Society Journals}

% make the titleuser-only
\maketitle

% \IEEEpeerreviewmaketitle

% \IEEEtitleabstractindextext{% 
\begin{abstract}
Growing concern about environmental sustainability (e.g., reducing carbon emissions and resource use) and public health has motivated ``green'' recommender systems that steer users toward more eco-friendly and healthier choices. However, many existing green recommendation approaches require training new models from scratch, incurring substantial computational and energy costs. Reranking-based methods, meanwhile, introduce an additional sorting stage at inference, increasing latency and computational cost.
In this work, we propose GRACE (Green Recommendation via Adaptive Conflict-rEsolution), a fine-tuning framework that integrates item-level sustainability signals (e.g., eco-scores or health indices) into pretrained recommendation models. Since these green values are usually discrete and non-differentiable, existing methods often rely on pairwise comparisons to promote greener items. GRACE instead introduces a differentiable approximation that enables direct optimization of the green criterion. To balance sustainability and personalization quality, GRACE further employs a gradient projection mechanism to mitigate conflicts between the green objective and the accuracy objective during fine-tuning. 
Experiments on real-world datasets demonstrate that GRACE improves sustainability-oriented recommendation outcomes while generally preserving recommendation accuracy through a controllable preference-anchored update mechanism.
\end{abstract}

% Note that keywords are not normally used for peer review papers.

% \IEEEraisesectionheading{\section{Introduction}\label{sec:introduction}}

\section{Introduction}
Recommender systems are now widely deployed across digital platforms and have reshaped consumer behavior by influencing what people buy, watch, and choose in their daily lives~\cite{rostami2023novel,yu2024self,raji2024commerce,wu2023survey}. 
As concerns about environmental sustainability and public health grow worldwide, these systems are increasingly viewed as a practical tool for nudging users toward more sustainable and healthier choices~\cite{arabzadeh2024green,beel2025green}.
Food is a particularly promising domain because food choices are frequent, highly actionable, and closely linked to both environmental impact and personal health~\cite{rostami2024novel,geng2023hybrid}. 
In this paper, we study green food recommendation, which aims to recommend foods that are healthier for users and have lower environmental impact, while still respecting individual preferences. We define ``green'' using item-level sustainability labels from external sources, such as eco-scores and health indices, which are often discrete ratings and can be used as supervision during training.

Food production and consumption play a critical role in both environmental and public health. The food system accounts for nearly 26\% of global greenhouse gas emissions~\cite{poore2018food}, and diet-related chronic diseases contribute to almost three-quarters of global mortality~\cite{bai2023global}. 
Encouraging modest shifts toward healthier, lower-impact diets can yield substantial benefits. Prior studies report potential emission reductions of up to 17\% globally and over 32\% in high-consumption populations~\cite{li2024diet}. Simple substitutions from high-carbon foods to lower-impact options can reduce dietary carbon footprints by over 35\% while improving nutrition~\cite{grummon2023diet}. These findings motivate integrating sustainability signals into personalized recommendation models so that everyday recommendations can produce meaningful gains for both people and the planet~\cite{zheng2023price,liu2023crossdomain}.

Green food recommendation naturally requires multi-objective optimization that maintains recommendation accuracy while improving the sustainability of recommended items~\cite{felfernig2023recommender,chang2023bundle}.
Existing methods are mainly retraining-based~\cite{jing2025bites} or reranking-based~\cite{kalisvaart2025towards,rostami2023towards}. Retraining-based approaches integrate sustainability into model training but often require training from scratch, leading to high computational costs~\cite{alshabanah2025meta,verachtert2023scheduling,zhu2025pckd} and potentially large carbon emissions~\cite{vente2024clicks}. Reranking-based approaches avoid retraining but often add an extra sorting step at inference, increasing latency and server-side computation for large candidate sets. They also typically treat sustainability as a post-processing constraint rather than a learned signal, which may generalize poorly in practice~\cite{felfernig2023recommender}.

A key challenge is that sustainability labels are often discrete and non-differentiable, making them hard to optimize directly with gradient-based recommenders~\cite{felfernig2023recommender}. As a result, many methods fall back to pairwise ranking objectives with negative sampling~\cite{9843966}. However, for discrete sustainability supervision, these objectives can yield weak or unstable training signals and may not align well with the target sustainability objective~\cite{lee2021differentiable,karmim2024item,liu2023bayesian,shi2024enhanced,li2023meta}.

Moreover, incorporating sustainability objectives can introduce a trade-off with personalization, since over-emphasizing sustainability may distort preference learning and reduce recommendation accuracy~\cite{beel2025green,zhou2024advancing,yu2024xsimgcl}. This is particularly important when sustainability is enforced through optimization, because the update direction that improves sustainability can interfere with the direction that preserves user preferences~\cite{yuan2021one,rostami2023towards}. For example, aggressively prioritizing low-carbon vegan meals may overlook users’ habitual preferences for meat-based or high-calorie dishes, reducing satisfaction and engagement.

To address these challenges, we formulate green recommendation as a preference-preserving fine-tuning problem over discrete top-K item-side sustainability signals. Different from standard recommendation fine-tuning, the auxiliary green supervision is list-dependent, often non-differentiable, and may conflict with user-preference optimization. This formulation motivates GRACE (Green Recommendation via Adaptive Conflict-rEsolution), an efficient fine-tuning framework that integrates green objectives into pretrained recommendation models while preserving personalization quality and avoiding expensive retraining. GRACE introduces a differentiable approximation of discrete sustainability signals using Gumbel-softmax relaxation, enabling direct gradient-based optimization toward green objectives. In addition, GRACE employs a gradient projection mechanism to mitigate conflicts between sustainability and personalization gradients, so that progress on one objective does not unduly compromise the other. Extensive experiments on real-world datasets show that GRACE improves the sustainability of recommendations while achieving a favorable sustainability-accuracy trade-off. Our main contributions are as follows:
\begin{itemize}
    \item We formulate sustainability-aware recommendation as a preference-preserving fine-tuning problem over discrete top-K green signals, and propose GRACE to integrate item-level sustainability supervision into pretrained recommender models without full retraining or inference-time reranking.
    \item We introduce a differentiable objective for discrete sustainability labels via Gumbel-softmax relaxation, enabling direct optimization rather than relying solely on pairwise ranking and negative sampling.
    \item We develop a gradient projection strategy to mitigate objective conflicts during fine-tuning, achieving a better balance between sustainability and recommendation accuracy than reranking baselines across real-world datasets.
    % \item GRACE utilizes a differentiable approximation to optimize discrete sustainability signals and incorporates a gradient projection mechanism, effectively balancing sustainability improvements with personalization accuracy.
    % \item By fine-tuning GRACE on existing recommendation systems, experiments demonstrate consistent improvements across most metrics, including both recommendation accuracy and greenness compared to pretrained recommendation models. GRACE also outperforms existing reranking methods in effectively balancing accuracy and sustainability objectives.
\end{itemize}

\section{Related Work}
\label{sec:rw}

\subsection{Green Recommendation}
In recent years, there has been growing research interest in incorporating sustainability objectives into recommender systems to encourage consumers to adopt more sustainable lifestyles. Beyond traditional goals of personalization accuracy, these systems explicitly consider environmental and health-related impacts with the aim to deliver benefits not only for individual users but also for the whole society. The food domain has emerged as a key focus in this research, as dietary choices are closely linked to chronic non-communicable diseases and contribute significantly to global greenhouse gas emissions~\cite{vos2022intervention,poore2018food}. 

Many research efforts have explored ways to incorporate sustainability objectives into recommender systems. The GRAPE model~\cite{jing2025bites} introduces customized green loss functions to jointly optimize user satisfaction along with multiple green indicators, including both health-related and environmental metrics. The Carbon Footprint-Aware Recommender System (CFARS)~\cite{kalisvaart2025towards} uses a reranking approach to include greenness metrics with minimal impact on accuracy. FHFRS~\cite{rostami2023towards} adds fairness constraints to ensure healthier recipe options are more equitably presented. Other methods build on reranking by incorporating environmental certifications~\cite{beel2025green} or applying behavioral nudging strategies~\cite{arabzadeh2024green}. Despite these advances, most existing approaches depend on post-hoc reranking or other non-differentiable adjustments, making it difficult to fully integrate sustainability objectives into the model's learning process. This limitation highlights the need for differentiable frameworks that can directly optimize for green goals alongside user preferences, enabling a more seamless and effective alignment of personalization with environmental and health considerations.

% To operationalize such objectives, several models have been proposed. The GRAPE model~\cite{jing2025bites} introduces customized green loss functions to jointly optimize user satisfaction and environmental objectives. The Carbon Footprint-Aware Recommender System (CFARS)~\cite{kalisvaart2025towards} adopts a reranking strategy that integrates greenness metrics with minimal accuracy trade-offs, while FHFRS~\cite{rostami2023towards} incorporates fairness constraints to ensure equitable exposure of healthier recipe options. Other approaches extend re-ranking by leveraging environmental certifications~\cite{beel2025green} or integrating behavioral nudging principles~\cite{arabzadeh2024green}.  

However, the development of green recommender systems has been limited by the lack of widely available benchmark datasets. Most existing datasets that include both user–item interactions and sustainability assessments are focused on the food domain, where environmental and health impacts are relatively easier to quantify. For example, \textit{GreenRec}~\cite{zhang2024greenrec} aligns user interactions with nutritional and environmental indicators, while \textit{RecipeEmission}~\cite{kalisvaart2025towards} estimates recipe-level carbon footprints using ingredient-level emission data. While these datasets provide valuable testbeds, they also reveal the narrow scope of current resources, which are largely limited to food-related items and rely on domain-specific estimation pipelines. In this work, we evaluate GRACE using \textit{GreenRec} and \textit{RecipeEmission}, but the proposed fine-tuning framework can be extended in principle to other domains once reliable item-level sustainability signals are available.

% Nevertheless, the development of sustainability-aware recommenders has been constrained by the limited availability of benchmark datasets. Existing resources are predominantly concentrated in the food domain, where environmental and health impacts are easier to quantify. For example, GreenRec~\cite{zhang2024greenrec} is constructed by aligning user–item interactions with nutritional and environmental indicators, while RecipeEmission~\cite{kalisvaart2025towards} estimates recipe-level carbon footprints from ingredient-level emission factors. These datasets provide valuable testbeds but also reveal the narrow scope of current resources, as they are restricted to food-related items and rely on domain-specific estimation pipelines. In this work, we adopt GreenRec and RecipeEmission as representative datasets for evaluation, while acknowledging the broader need for sustainability-oriented datasets beyond the food sector.  

% Despite these advances, most existing methods rely on post-hoc reranking or other non-differentiable adjustments, which limit their ability to seamlessly incorporate sustainability goals into the optimization process. This gap motivates the exploration of differentiable frameworks that can directly embed green objectives into the learning procedure, thereby harmonizing user preference optimization with environmental and health considerations. 

\begin{figure*}[t] 
\centering 
% \vspace{-3mm} 
\includegraphics[width=.8\textwidth]{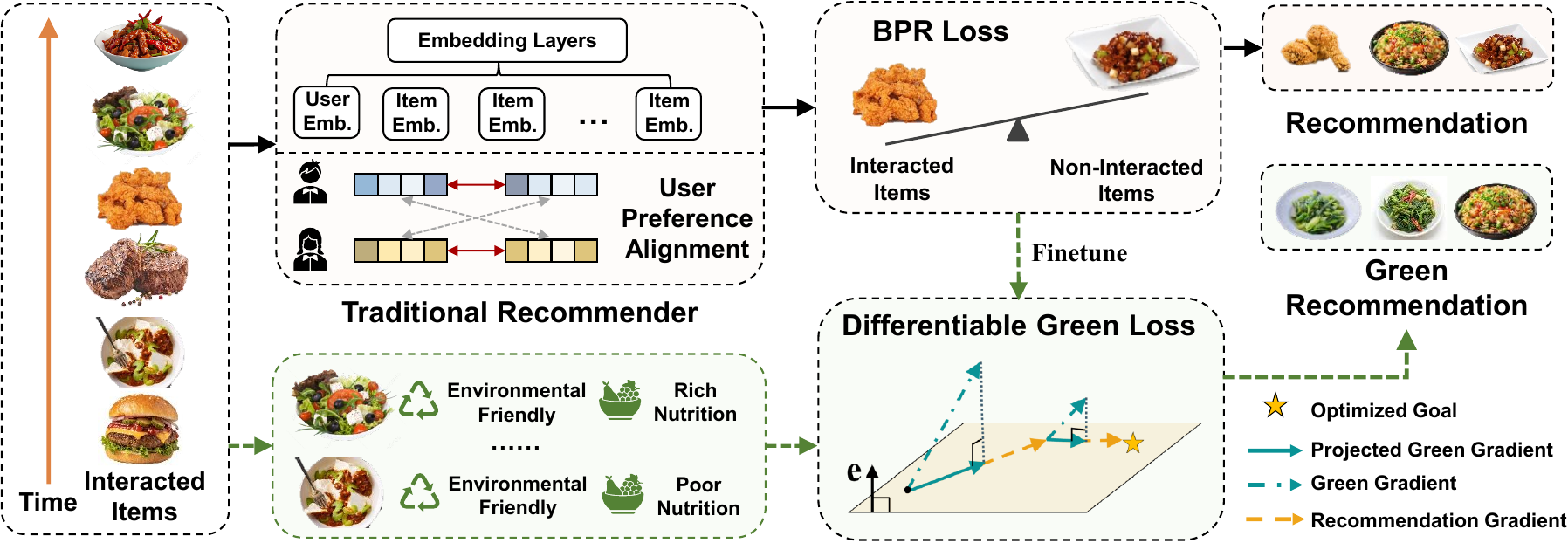} 
% \vspace{-10mm} 
\caption{The GRACE framework. GRACE fine-tunes pretrained recommendation models by jointly optimizing for accuracy and sustainability. It employs a differentiable green loss and a gradient projection mechanism to balance conflicting objectives, enabling greener recommendations while preserving accuracy.
% \textcolor{red}{For existing recommender systems, GRACE integrates sustainability objectives through differentiable ranking of top-$K$ metrics, leverages soft permutation matrices to jointly optimize green and preference signals, and finally applies gradient projection with norm control to fine-tune pre-trained models while preserving recommendation accuracy.}
} 
\label{fig:framework} 
% \vspace{-5mm} 
\end{figure*}

\subsection{Differentiable Optimization for Recommender Systems}

A key challenge in recommender systems is that ranking-based evaluation metrics are often discrete and non-differentiable, making them difficult to optimize using standard gradient-based methods. To address this, early approaches like Bayesian Personalized Ranking (BPR)~\cite{rendle2009bpr} reformulate the task as a pairwise preference learning problem. While BPR is widely used and often performs well, its effectiveness heavily depends on the quality of negative sampling. To improve on this, researchers have proposed various extensions, such as smoother approximations~\cite{liang2018top}, improved posterior estimation~\cite{wang2023setrank}, and causal modeling approaches~\cite{zhao2024causal}. Despite these efforts, BPR and its variants still face challenges due to misalignment between optimization directions and evaluation goals.

% A key challenge in recommender systems arises from the discrete and non-differentiable nature of ranking-based evaluation metrics, which prevents direct gradient-based learning. To circumvent this issue, early solutions such as Bayesian Personalized Ranking (BPR)~\cite{rendle2009bpr} reformulated the problem as pairwise preference learning. While effective, the performance of BPR critically depends on the quality of negative sampling, and various improvements have been proposed, including smoother approximation~\cite{liang2018top}, refined posterior estimation~\cite{wang2023setrank}, and causal modeling perspectives~\cite{zhao2024causal}. However, BPR and its variants still suffer from misalignment between optimization directions and evaluation goals.

To address the challenge of non-differentiable ranking metrics, recent work has introduced differentiable ranking approaches that relax discrete operations into continuous ones. NeuralSort~\cite{grover2018stochastic} is one of the pioneering methods to make the sorting operation differentiable, enabling gradient-based optimization of ranking objectives. Building on this, Pobrotyn and Białobrzeski~\cite{pobrotyn2021neuralndcg} extend the idea to directly optimize NDCG, which inspired further methods such as DRM~\cite{lee2021differentiable} and diffNDCG~\cite{zhou2024optimizing}. More recently, Differentiable Hit (DH), proposed in FADE~\cite{yoo2024ensuring}, uses the Gumbel-softmax trick to compute stable and efficient gradients for discrete hit ratio metrics. Other efforts have also explored differentiable approximations for Top-$k$ retrieval~\cite{blondel2020fast} and rank aggregation~\cite{prillo2020softsort}, expanding the use of continuous optimization in recommendation and information retrieval tasks.

% Differentiable ranking approaches have emerged that relax discrete operations into continuous ones. NeuralSort~\cite{grover2018stochastic} pioneered this line by introducing a differentiable relaxation of the sorting operator. Pobrotyn and Białobrzeski~\cite{pobrotyn2021neuralndcg} extended this idea to directly optimize NDCG, inspiring further work such as DRM~\cite{lee2021differentiable} and diffNDCG~\cite{zhou2024optimizing}. More recently, Differentiable Hit (DH), proposed in FADE~\cite{yoo2024ensuring}, leverages the Gumbel-Softmax trick to enable stable and efficient gradient computation for discrete hit ratios. Parallel efforts also explore differentiable Top-$k$ approximations for retrieval~\cite{blondel2020fast} and rank aggregation~\cite{prillo2020softsort}, broadening the scope of continuous optimization techniques in information retrieval and recommendation. 

Nevertheless, to the best of our knowledge, differentiable optimization frameworks have not yet been applied to sustainability-aware recommendation. Existing methods either do not effectively handle gradient conflicts in multi-objective optimization or rely on non-differentiable ways to incorporate green metrics. This gap motivates the design of new approaches that embed sustainability objectives into differentiable ranking frameworks, allowing user preferences to be optimized alongside environmental and health goals in a unified and efficient manner.

% Nevertheless, to the best of our knowledge, differentiable optimization frameworks are not applied in sustainability-aware recommendation. Existing methods either inadequately resolve gradient conflicts in multi-objective optimization or fail to integrate green metrics in a fully differentiable manner. This gap motivates the design of new approaches that embed sustainability objectives into differentiable ranking frameworks, harmonizing user preference optimization with environmental and health goals.

\subsection{Multi-objective Optimization for Recommender Systems}

Real-world recommender systems often need to balance recommendation accuracy with additional objectives, such as revenue, conversion, diversity, fairness, user satisfaction, and long-term engagement. A common solution is to formulate recommendation as a multi-objective optimization problem and search for a trade-off among competing goals. For example, Pareto-Efficient Recommendation~\cite{lin2019pareto} studies the joint optimization of GMV and CTR in e-commerce recommendation, while MGDRec~\cite{milojkovic2020multi} introduces multi-gradient descent to combine recommendation objectives with different scales. Following this line, PAPERec~\cite{xie2021personalized} learns personalized Pareto-efficient trade-offs, enabling different users to have different preferences over multiple objectives.

Recent studies further extend multi-objective recommendation from global trade-off modeling to adaptive optimization strategies. PMORS~\cite{jin2024pareto} incorporates a forgetting-curve mechanism into Pareto optimization to balance negative feedback and recommendation utility in short-video recommendation. MultiTRON~\cite{wilm2024pareto} approximates the Pareto front for session-based recommendation by conditioning the model on different preference vectors. GradCraft~\cite{bai2024gradcraft} focuses on multi-task recommendation and mitigates task interference by jointly considering gradient magnitude and gradient direction. These methods show that explicitly modeling objective conflicts is important for recommendation systems with multiple practical goals.

These studies provide useful references for handling competing optimization signals in recommender systems. Our work follows this general motivation and studies green fine-tuning for pretrained recommenders, where sustainability supervision is introduced after the model has learned user preferences. In this setting, improving sustainability requires controlling the influence of the green objective on the original preference-learning direction. GRACE addresses this requirement with an asymmetric update strategy, where the recommendation gradient serves as the anchor and the sustainability gradient is projected and constrained before model updating.

\section{Methodology}
\label{sec:method}

% \subsection{Problem Definition}
% As recommender systems play a growing role in shaping user choices, there is increasing interest in making them more sustainable by encouraging selections with lower environmental impact or better health value. However, incorporating such objectives without sacrificing personalization or requiring full model retraining remains a significant challenge.

% Specifically, we consider fine-tuning a pre-trained recommendation model $f(\theta)$, which outputs raw prediction scores $\mathbf{s}\in\mathbb{R}^V$ over a set of $V$ candidate items. Each item is associated with green attributes (e.g., carbon footprint, health index), which can be incorporated into the training objective through corresponding scores. 
% The goal is to fine-tune $f(\theta)$ so that the resulting ranked lists maintain high accuracy while also improving the sustainability of recommended items. 

\subsection{Problem Definition}

In this work, we study green food recommendation, which aims to improve the environmental and health-related indices of recommended items while preserving personalization accuracy. More specifically, we use ``green'' to mean item-side sustainability signals that describe an item’s environmental and/or health desirability independent of the user’s preferences. Following \textit{GreenRec}~\cite{zhang2024greenrec} and \textit{RecipeEmission}~\cite{kalisvaart2025towards}, these signals are provided as external annotations. \textit{GreenRec} labels each recipe with three indicators, namely an Environmental Impact Score that aggregates multiple environmental factors such as greenhouse-gas emissions, land use, water stress, and eutrophication, as well as two health-related indicators, the Nutritional Impact Score and the Healthy Meal Index~\cite{zhang2024greenrec}. \textit{RecipeEmission} provides an explicit environmental measure through the item-level CO$_2$-eq footprint per recipe, and derives a bounded greenness score from CO$_2$ such that higher greenness indicates lower emissions~\cite{kalisvaart2025towards}. Overall, we treat green signals as externally measurable item properties that can be optimized jointly with preference supervision.

Formally, we fine-tune a pretrained recommender model $f(\theta)$ that outputs predicted scores $\mathbf{s}\in\mathbb{R}^V$ over $V$ candidate items. Each item $v$ is associated with one or more green attributes (e.g., CO$_2$-eq, greenness, or health indices) as defined above. Our objective is to fine-tune $f(\theta)$ so that the resulting ranked lists achieve high recommendation accuracy while increasing the sustainability of the recommended items.

\subsection{GRACE}
As shown in Figure~\ref{fig:framework}, we propose GRACE, a training-time fine-tuning framework that transforms a standard pretrained recommender into a green recommender without retraining from scratch and without changing the inference pipeline. Compared with traditional recommenders that optimize only user preference signals, GRACE incorporates item-level sustainability labels during fine-tuning. Unlike reranking-based methods that enforce sustainability after scoring and introduce additional inference-time sorting, GRACE integrates sustainability in the model parameters so that greener ranked lists are produced directly by the base model.

% Concretely, GRACE introduces two key components:
% 1) a Differentiable Green Loss that enables direct gradient-based optimization on discrete green labels, and 2) a gradient projection mechanism that reduces conflicts between sustainability and accuracy updates. Together, these components improve recommendation sustainability while preserving personalization accuracy.\footnote{The code is available at \url{https://anonymous.4open.science/r/GRACE-7E32}.} For clarity, Table~\ref{tab:notation} summarizes the key notations used throughout the methodology.

Concretely, GRACE introduces two key components:
1) a Differentiable Green Loss that enables direct gradient-based optimization on discrete green labels, and 2) a gradient projection mechanism that reduces conflicts between sustainability and accuracy updates. Together, these components improve recommendation sustainability while preserving personalization accuracy. For clarity, Table~\ref{tab:notation} summarizes the key notations used throughout the methodology.

\begin{table}[t]
\centering
\caption{Key Notations Used in GRACE.}
\label{tab:notation}
\footnotesize
\resizebox{\linewidth}{!}{
\begin{tabular}{cl}
\toprule
Notation & Description \\
\midrule
$f(\theta)$ & Pretrained recommendation model with parameters $\theta$ \\
$\mathbf{s}\in\mathbb{R}^{V}$ & Predicted scores over $V$ candidate items \\
$s_i$, $\hat{s}_i$, $\tilde{s}_i$ & Raw, normalized, and perturbed score of item $i$ \\
$\tau$ & Temperature parameter for soft ranking \\
$N$ & Top-ranked positions in differentiable optimization \\
$K$ & Cutoff length used for evaluation metrics \\
$\hat{\mathbf{P}}$ & Soft permutation matrix for differentiable ranking \\
$\mathbf{y}\in\{0,1\}^{V}$ & Ground-truth relevance vector \\
$DH(n)$ & Differentiable hit indicator at rank position $n$ \\
$\mathrm{DGS}(n)$ & Differentiable green score at rank position $n$ \\
$e_i^r$ & Scalar sustainability score of item $i$ used for training \\
$L_{\mathrm{green}}$ & Differentiable sustainability loss \\
$L_{\mathrm{rec}}$ & Differentiable recommendation loss \\
$\mathbf{g}_{\mathrm{green}}$ & Gradient of $L_{\mathrm{green}}$ with respect to $\theta$ \\
$\mathbf{g}_{\mathrm{rec}}$ & Gradient of $L_{\mathrm{rec}}$ with respect to $\theta$ \\
$\mathbf{g}_{\mathrm{proj}}$ & Projected sustainability gradient \\
$\mathbf{g}_{\mathrm{final}}$ & Final update direction for fine-tuning \\
$\alpha$ & Projection ratio for green updates \\
\bottomrule
\end{tabular}
}
\end{table}

\subsubsection{Differentiable Approximation of Green Objectives}

Green recommendation evaluates sustainability on the top-K items presented to a user. However, generating this list requires sorting and top-K selection, which are non-differentiable. Moreover, sustainability supervision is usually item-level and discrete, provided as labels or rating levels from external sources. As a result, sustainability objectives cannot be directly backpropagated through ranking, making joint optimization of sustainability and accuracy challenging with standard gradient-based training.

To address this issue, GRACE adopts a differentiable relaxation of list ranking based on Gumbel-softmax perturbations, a general differentiable sorting technique that has been explored in prior work on learning listwise objectives (e.g.,~\cite{yoo2024ensuring}). Instead of treating sustainability as a post-processing constraint or relying solely on pairwise surrogates, GRACE optimizes a list-level sustainability objective through a smooth approximation of the top-$K$ operator, reducing computational overhead and ensuring bounded gradients during fine-tuning. This design is well suited to green recommendation, as sustainability objectives explicitly depend on which items appear near the top of the ranked list.

\noindent\textbf{Score Normalization.} To ensure stable gradient computation and prevent numerical instability, such as gradient explosions during optimization, we first normalize raw prediction scores $\mathbf{s}\in\mathbb{R}^V$ (where $V$ is the number of items) using $L_2$ normalization:
\begin{equation}
\hat{s}_i = \frac{s_i}{\|\mathbf{s}\|_2 + \epsilon},
\end{equation}
where $\hat{s}_i$ denotes the normalized score of item $i$ and $\epsilon>0$ prevents division by zero.

\noindent\textbf{Differentiable Perturbation with Gumbel Noise.} To approximate discrete selection while retaining differentiability, we inject standard Gumbel noise into the normalized scores:
\begin{equation}
\tilde{s}_i = \hat{s}_i + g_i,
\end{equation}
where $g_i$ is sampled as
\begin{equation}
g_i = -\log(-\log(U_i)), \quad U_i \sim \text{Uniform}(0,1),
\end{equation}
so that $g_i$ follows the standard Gumbel $(0,1)$ distribution. This perturbation induces a smooth, stochastic relaxation of ranking decisions. Intuitively, it defines a distribution over rankings that is amenable to gradient-based optimization.

\noindent\textbf{Efficient Soft Permutation Matrix Construction.} Traditional soft permutation matrix construction often relies on all pairwise score differences, which requires forming a $V\times V$ matrix and costs $O(V^2)$. This is prohibitive when $V$ is large. GRACE instead uses a NeuralSort-style reformulation that avoids materializing pairwise differences and computes only the top-$N$ rows, which are sufficient for top-$N$ objectives.

Given perturbed scores $\tilde{\mathbf{s}}$, we compute the rank $\mathrm{rank}(\tilde{s}_i)$ of each item and the prefix sum $\sum_{j:\tilde{s}_j < \tilde{s}_i} \tilde{s}_j$, both efficiently obtainable via sorting and prefix-sum operations. The approximation is defined as:
\begin{equation}
\hat{A}_{i} = (2\text{rank}(\tilde{s}_i) - V) \tilde{s}_i - 2 \sum_{j:\tilde{s}_j < \tilde{s}_i} \tilde{s}_j,
\label{eq:score}
\end{equation}
where the first term measures relative position scaled by score and the second term efficiently aggregates smaller-score influences. The soft permutation probabilities for ranks $i=1,\dots,N$ are then computed as
\begin{equation}
\begin{split}
\hat{P}_{i,:} = \text{softmax} \left( \frac{(V + 1 - 2i) \tilde{\mathbf{s}} - \hat{A}}{\tau} \right), \quad
i=1, \dots, N,
\end{split}
\label{eq:topk}
\end{equation}
where $\tau>0$ is the temperature that controls smoothness. Compared with full pairwise sorting methods that incur $O(V^2)$ complexity, GRACE sorts $\tilde{\mathbf{s}}$ in $O(V\log V)$ time and computes prefix sums in $O(V)$ time. Constructing each of the $N$ softmax rows requires $O(V)$ time, resulting in an overall per-instance complexity of $O(V\log V + NV)$.

\noindent\textbf{Differentiable Rank-wise Signals.}
Utilizing the soft permutation matrix, GRACE constructs two differentiable rank-wise signals for recommendation accuracy and sustainability optimization, respectively. Let $\mathbf{y}\in\{0,1\}^{V}$ denote the ground-truth relevance vector, where $y_i=1$ indicates that item $i$ is relevant. Following the idea of Differentiable Hit, we define the soft relevance indicator at rank position $n$ as
\begin{equation}
\mathrm{DH}(n)=\sum_{i=1}^{V}\hat{P}_{n,i}y_i .
\end{equation}
This quantity lies in $[0,1]$ and provides a differentiable proxy for the discrete hit indicator at rank position $n$.

For sustainability optimization, GRACE replaces the relevance indicator with the scalar sustainability score. Let $e_i^r$ denote the scalar sustainability score of item $i$ used for training. We define the differentiable green score at rank position $n$ as
\begin{equation}
\mathrm{DGS}(n)=\sum_{i=1}^{V}\hat{P}_{n,i}e_i^r .
\end{equation}
This quantity represents the expected sustainability score of the item placed at rank position $n$ under the relaxed ranking distribution. Therefore, the same soft permutation matrix provides both a relevance-aware signal for recommendation accuracy and a sustainability-aware signal for green optimization. The following proposition formalizes that these relaxed rank-wise signals are consistent with their discrete counterparts.

\begin{theorem}
Consistency of Differentiable Rank-wise Signals.
Let $\hat{\mathbf{P}}(\tau)$ be defined in Eq.~(\ref{eq:topk}), and let $\mathbf{P}$ denote the discrete permutation matrix produced by sorting the perturbed scores $\tilde{\mathbf{s}}$. Under Gumbel perturbation, all $\tilde{s}_i$ are distinct with probability one. For any rank position $n$,
\begin{equation}
\lim_{\tau\to 0}\mathrm{DH}(n;\tau)
=
\sum_{i=1}^{V}P_{n,i}y_i ,
\end{equation}
and
\begin{equation}
\lim_{\tau\to 0}\mathrm{DGS}(n;\tau)
=
\sum_{i=1}^{V}P_{n,i}e_i^r .
\end{equation}

Proof.
In Eq.~(\ref{eq:topk}), the $n$-th row $\hat{\mathbf{P}}_{n,\cdot}(\tau)$ is obtained by applying a softmax with temperature $\tau$ to a vector of logits determined by the perturbed scores $\tilde{\mathbf{s}}$. Under Gumbel perturbation, the entries of $\tilde{\mathbf{s}}$ are distinct with probability one, so the item placed at the $n$-th rank is unique. Let $i^\star$ denote its index. By construction, the corresponding logit in row $n$ is the unique maximum, and scaling the logits by $1/\tau$ does not change the maximizer.

A standard property of softmax states that if a vector has a unique maximum at $i^\star$, then as $\tau\to 0$, the softmax converges to the one-hot vector at $i^\star$. Hence,
\begin{equation}
\lim_{\tau \to 0} \hat{P}_{n,i^\star}(\tau)=1,
\qquad
\lim_{\tau \to 0} \hat{P}_{n,i}(\tau)=0 \;\; (i\neq i^\star).
\end{equation}
Therefore,
\begin{equation}
\begin{split}
\lim_{\tau\to 0}\mathrm{DH}(n;\tau)
&=
\sum_{i=1}^{V}
\left(
\lim_{\tau\to 0}\hat{P}_{n,i}(\tau)
\right)y_i  \\
&=
\sum_{i=1}^{V}P_{n,i}y_i .
\end{split}
\end{equation}
and similarly,
\begin{equation}
\begin{split}
\lim_{\tau\to 0}\mathrm{DGS}(n;\tau)
&=
\sum_{i=1}^{V}
\left(
\lim_{\tau\to 0}\hat{P}_{n,i}(\tau)
\right)e_i^r  \\
&=
\sum_{i=1}^{V}P_{n,i}e_i^r .
\end{split}
\end{equation}
Here, the second equality follows because the limiting soft permutation row is exactly the $n$-th row of the discrete permutation matrix $\mathbf{P}$. This completes the proof.
\end{theorem}

\noindent\textbf{Differentiable Green Objective.}
GRACE defines a differentiable sustainability loss $L_{\mathrm{green}}(\theta)$ by maximizing the expected average sustainability score over the relaxed top-$N$ list:
\begin{equation}
\begin{split}
L_{\mathrm{green}}(\theta)
&=
-\frac{1}{B}
\sum_{b=1}^{B}
\frac{1}{N}
\sum_{n=1}^{N}
\mathrm{DGS}^{(b)}(n) \\
&=
-\frac{1}{B}
\sum_{b=1}^{B}
\frac{1}{N}
\sum_{n=1}^{N}
\sum_{i=1}^{V}
\hat{P}^{(b)}_{n,i} e_i^r .
\end{split}
\label{eq:lgreen}
\end{equation}
where $B$ denotes the batch size, $\hat{P}^{(b)}_{n,i}$ denotes the soft probability that item $i$ is placed at rank position $n$ for user instance $b$, and $e_i^r$ denotes the scalar sustainability score of item $i$. The factor $1/N$ averages the expected sustainability score over the relaxed top-$N$ list. This objective encourages items with higher sustainability scores to receive larger soft exposure in the top-ranked positions, thereby enabling end-to-end optimization of list-level sustainability.

\noindent\textbf{Differentiable Accuracy Objective.}
To maintain recommendation quality, GRACE uses a differentiable approximation of NDCG@$N$ as the accuracy objective. Given the ideal DCG $\mathrm{maxDCG}^{(b)}$ for user instance $b$, we define
\begin{equation}
L_{\mathrm{rec}}(\theta)
=
-\frac{1}{B}
\sum_{b=1}^{B}
\frac{1}{\mathrm{maxDCG}^{(b)}}
\sum_{n=1}^{N}
\frac{\mathrm{DH}^{(b)}(n)}{\log_2(n+1)} .
\end{equation}
Together, $L_{\mathrm{green}}$ and $L_{\mathrm{rec}}$ form a unified differentiable objective for multi-objective fine-tuning.

To motivate score normalization, we show that it yields bounded gradients for both losses.
% To justify the use of score normalization and to explain the observed training stability, we further show that the gradients of both losses remain bounded.

\begin{theorem}
Bounded Gradients.
Assume $\|\hat{\mathbf{s}}\|_2\le 1$ and $|e_i^r|\le E_{\max}$ for all items. Then the gradients of $L_{\mathrm{green}}$ and $L_{\mathrm{rec}}$ with respect to $\tilde{\mathbf{s}}$ are bounded:
\begin{equation}
\|\nabla_{\tilde{\mathbf{s}}}L_{\mathrm{green}}\|_2\le C_{\mathrm{green}},
\qquad
\|\nabla_{\tilde{\mathbf{s}}}L_{\mathrm{rec}}\|_2\le C_{\mathrm{rec}},
\end{equation}
for constants depending only on $(B,N,V,\tau,E_{\max})$.

Proof.
Both losses are weighted sums of entries of the soft permutation matrix $\hat{\mathbf{P}}$. Specifically, $L_{\mathrm{rec}}$ weights $\hat{P}_{n,i}$ by the bounded relevance indicator $y_i$, while $L_{\mathrm{green}}$ weights $\hat{P}_{n,i}$ by the bounded sustainability score $e_i^r$. Therefore, it suffices to bound $\nabla_{\tilde{\mathbf{s}}}\hat{P}_{n,i}$.

The soft permutation matrix in Eq.~(\ref{eq:topk}) is produced by applying softmax to logits determined by the perturbed scores. Under Gumbel perturbation, ties occur with probability zero. Thus, within each fixed ordering region, the rank and prefix-sum structure is fixed, and the logits are affine functions of $\tilde{\mathbf{s}}$. Their Jacobian is therefore bounded by a constant depending only on $(V,\tau)$. Since the softmax mapping has a bounded Jacobian, each $\nabla_{\tilde{\mathbf{s}}}\hat{P}_{n,i}$ is bounded almost everywhere.

Because $y_i\in\{0,1\}$ and $|e_i^r|\le E_{\max}$, the coefficients in both $L_{\mathrm{rec}}$ and $L_{\mathrm{green}}$ are bounded. Since each loss is a finite weighted sum over $B$ user instances, $N$ rank positions, and $V$ candidate items, the full gradients are bounded by constants depending only on $(B,N,V,\tau,E_{\max})$.
\end{theorem}

We formulate fully differentiable listwise objectives for both sustainability and recommendation accuracy. The proposed relaxation recovers discrete ranking in the low-temperature limit, scales efficiently to large candidate sets, and yields bounded gradients under score normalization. Together, these properties support stable sustainability-aware fine-tuning, which we further combine with gradient projection to mitigate conflicts between objectives.

\subsubsection{Gradient Projection}
% Mitigating Catastrophic Forgetting via 
\label{sec:proj}

A key challenge in incorporating sustainability objectives into recommender systems is maintaining personalization quality during fine-tuning, since emphasizing sustainability can inadvertently shift the model away from user preference signals and reduce accuracy. This arises because the sustainability objective and the preference objective may induce different gradient directions, simply combining their updates can overemphasize one objective and degrade the other. To better balance these objectives, GRACE introduces a gradient projection mechanism. GRACE computes the preference gradient and the sustainability gradient separately, then projects the sustainability gradient to remove the component that aligns with the preference gradient’s direction. This yields updates that improve sustainability while minimally interfering with preference optimization, helping preserve recommendation accuracy during fine-tuning.

Specifically, given a pretrained recommendation model with parameters $\theta$, we compute the recommendation gradient $\mathbf{g}_{\text{rec}}$ and the sustainability gradient $\mathbf{g}_{\text{green}}$ separately as:
\begin{align}
\mathbf{g}_{\text{rec}} &= \nabla_{\theta} L_{\text{rec}}(\theta), \\
\mathbf{g}_{\text{green}} &= \nabla_{\theta} L_{\text{green}}(\theta),
\label{eq:grad}
\end{align}
where $\nabla_{\theta}$ denotes the gradient operator with respect to model parameters $\theta$. Specifically, $ \nabla_{\theta} L(\theta)$ yields a $P$-dimensional gradient vector, where each component corresponds to the partial derivative of $L(\theta)$ with respect to one parameter dimension. 

To prevent the sustainability gradient from adversely affecting recommendation quality, the sustainability gradient is projected onto the orthogonal complement of the recommendation gradient direction. The projected sustainability gradient $\mathbf{g}_{\text{proj}}$ is computed as:
\begin{equation}
\mathbf{g}_{\text{proj}} = \mathbf{g}_{\text{green}} - \frac{\mathbf{g}_{\text{green}}^\top \mathbf{g}_{\text{rec}}}{\|\mathbf{g}_{\text{rec}}\|_2^2 + \epsilon}\mathbf{g}_{\text{rec}}, \quad \epsilon=10^{-8},
\label{eq:project}
\end{equation}
where $\mathbf{g}_{\text{green}}^\top \mathbf{g}_{\text{rec}}$ denotes the inner product between the sustainability and recommendation gradients; $\|\mathbf{g}_{\text{rec}}\|_2$ represents the $L_2$ norm (Euclidean norm) of $\mathbf{g}_{\text{rec}}$, ensuring proper normalization of the projection scalar. This projection removes the component of the sustainability gradient along the recommendation gradient direction, yielding an orthogonal residual update that reduces first-order interference with preference optimization.

To further regulate the influence of sustainability during fine-tuning, GRACE applies a gradient norm constraint that bounds the magnitude of sustainability updates relative to preference updates. The final update direction is defined as:
\begin{equation}
\mathbf{g}_{\text{final}} = \mathbf{g}_{\text{rec}} + \min\left(1, \frac{\alpha \|\mathbf{g}_{\text{rec}}\|_2}{\|\mathbf{g}_{\text{proj}}\|_2 + \epsilon}\right)\mathbf{g}_{\text{proj}},
\label{eq:finalgrad}
\end{equation}
where $\alpha \in [0,1]$ is the projection ratio that caps the strength of sustainability updates to protect recommendation quality, $\|\mathbf{g}_{\text{proj}}\|_2$ is the $L_2$ norm of the projected sustainability gradient, and the $\min$ operator ensures the sustainability update magnitude does not exceed $\alpha$ times the recommendation gradient magnitude.
The gradient norm constraint controls the trade-off between sustainability and accuracy, balancing environmental and health gains with preference preservation.

% is a hyperparameter termed the projection ratio, designed to restrict the relative magnitude of sustainability updates, thereby safeguarding recommendation quality; $\|\mathbf{g}_{\text{proj}}\|_2$ is the $L_2$ norm of the projected sustainability gradient; the $\min$ operation ensures that the contribution of sustainability updates does not exceed a fraction $\alpha$ of the recommendation gradient’s magnitude. This carefully calibrated projection approach ensures incremental sustainability enhancement while rigorously protecting user preference models from deterioration.

% where $\alpha \in [0,1]$ is a hyperparameter termed the projection ratio, designed to restrict the relative magnitude of sustainability updates, thereby safeguarding recommendation quality; $\|\mathbf{g}_{\text{proj}}\|_2$ is the $L_2$ norm of the projected sustainability gradient; the $\min$ operation ensures that the contribution of sustainability updates does not exceed a fraction $\alpha$ of the recommendation gradient’s magnitude. This carefully calibrated projection approach ensures incremental sustainability enhancement while rigorously protecting user preference models from deterioration.

Subsequently, model parameters $\theta$ are updated via gradient descent using the combined gradient as:
\begin{equation}
\theta \leftarrow \theta - \eta \cdot \mathbf{g}_{\text{final}},
\label{eq:update}
\end{equation}
where $\eta$ denotes the learning rate. 

\subsection{Inference-Time Complexity Analysis}
Recommender systems generally follow two inference paradigms. Let $d$ denote the embedding dimension and $L$ the length of the user interaction sequence. In retrieval-based models, each user is encoded into a $d$-dimensional representation and matched against all $V$ item embeddings, resulting in a dominant inference cost of $O(Vd)$. In predictive scoring models, the user’s interaction sequence of length $L$ is first encoded with complexity depending on the architecture (e.g., $O(Ld^2)$ or $O(L^2 d)$ for attention-based encoders), followed by scoring all $V$ items with an additional $O(Vd)$ cost. We denote the overall inference complexity of the base recommender model generically as $R$.

\noindent\textbf{Traditional reranking based green recommendation.}
Reranking based approaches first compute item scores using the base recommender and then apply sustainability-aware reranking as a post-processing step. This introduces an explicit sorting operation over $n$ candidate items, which incurs $O(n\log n)$ time complexity. When reranking is applied to a large candidate set, with $n$ approaching the full item catalog size $V$, the additional cost becomes $O(V\log V)$. For retrieval-based models with $R = O(Vd)$, this sorting step adds a non-negligible logarithmic overhead on top of similarity computation. For predictive scoring models with $R = O(Ld^2 + Vd)$, the reranking cost dominates when $V\log V \gtrsim Ld^2 + Vd$, a common scenario in large-scale recommendation settings. Consequently, the inference-time complexity of reranking-based methods can be expressed as $R + O(n\log n)$ and approaches $R + O(V\log V)$ for large candidate sets, leading to increased inference latency.

\noindent\textbf{GRACE.}
In contrast, GRACE integrates sustainability objectives entirely during training and does not modify the inference pipeline of the base recommender. At serving time, the model performs user encoding, item scoring, and top-$n$ selection using the same operations as the original recommendation system, without introducing any additional reranking or sorting steps. As a result, the inference-time complexity of GRACE remains unchanged and can be written as $R_{\mathrm{GRACE}} = R$. By avoiding the explicit $O(n\log n)$ reranking overhead, GRACE maintains the same inference efficiency as the underlying recommender, which is especially important in large-catalog and low-latency deployment scenarios.

\begin{figure}[t!]
% \FloatBarrier
\centering
\captionsetup[subfloat]{}
    \subfloat[GreenRec]{
    \includegraphics[width=0.62\linewidth]{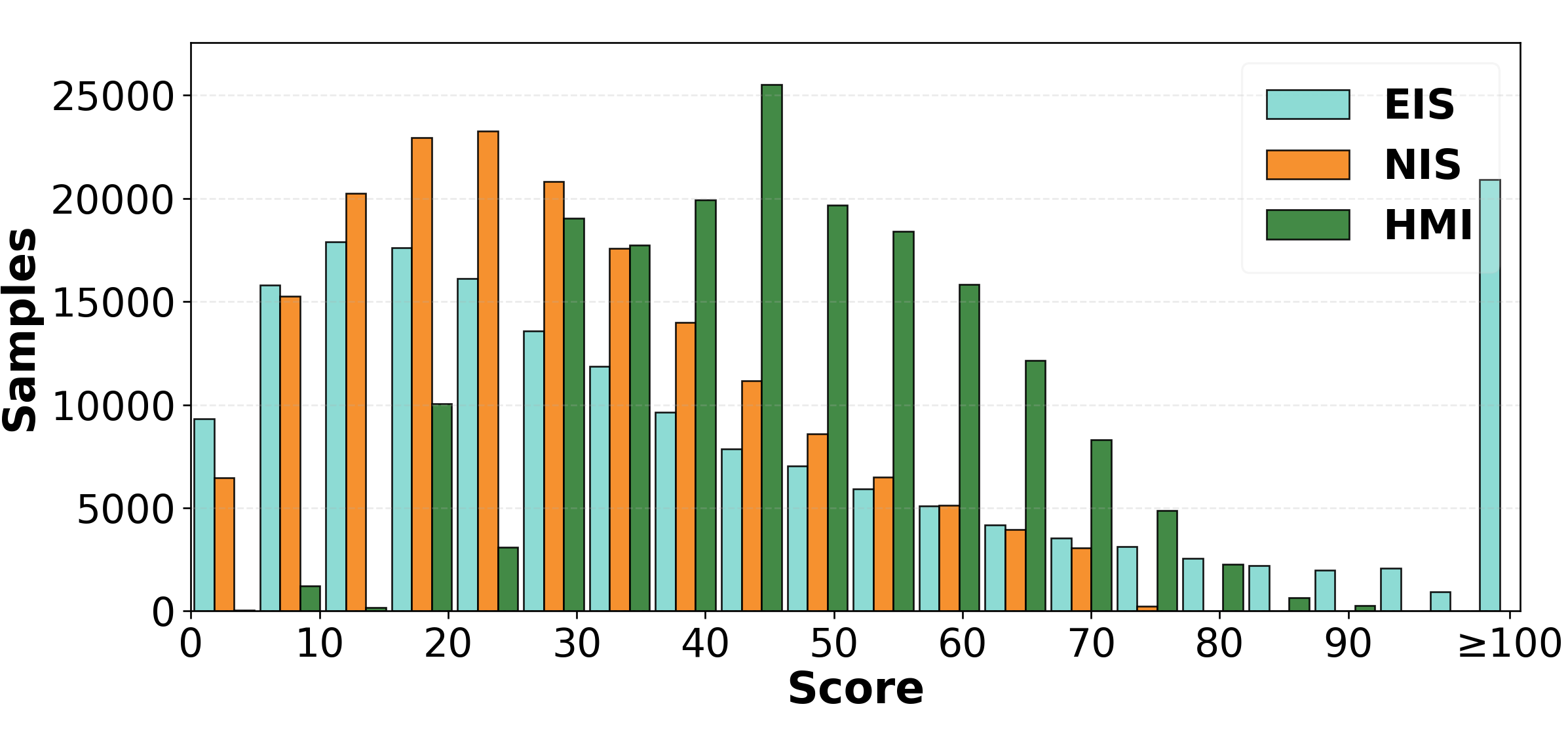}
    }
    \subfloat[RecipeEmission]{
    \includegraphics[width=0.3\linewidth]{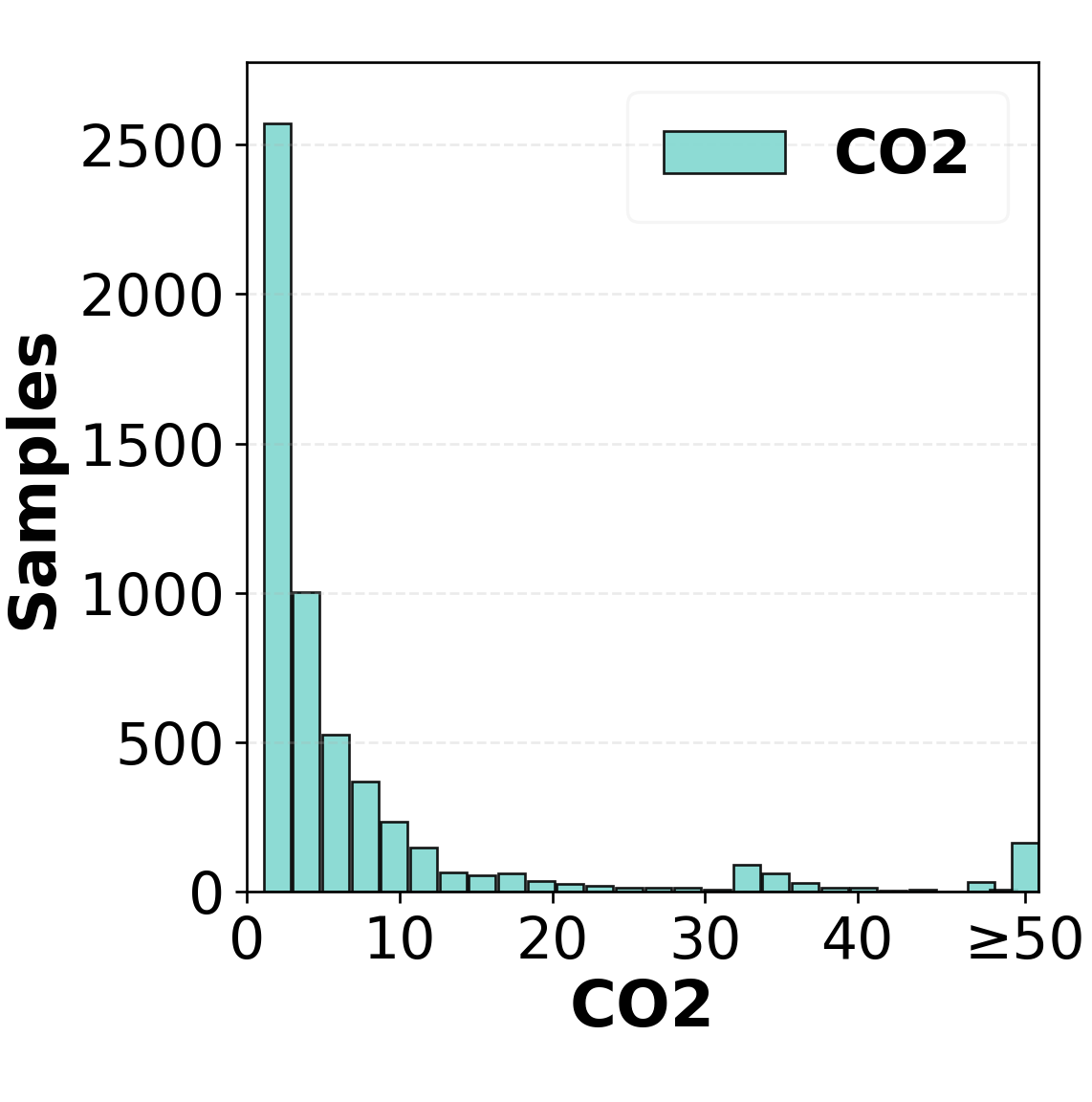}
    }
    % \vspace{-1mm}
    \caption{Distribution of Sustainability Indicators in the Two Datasets.}
    \label{fig:data}
    % \vspace{-3mm}
\end{figure}

\begin{table*}[h!]
\caption{
Performance comparison on \textit{GreenRec}. “Rel.” denotes GRACE’s relative improvement over the corresponding pretrained model. Best and second-best results are in bold and underlined, respectively. Cells are color-coded relative to the original pretrained model (green: improved, red: degraded, blank: unchanged).
% Performance Comparison with Baseline methods of Pretrained Recommendation Models on GreenRec Dataset. “Rel.” denotes the relative improvement achieved by GRACE over the original pretrained recommendation model. The best results are bold, and the second-best are underlined. Cells are color-coded by comparing each method with the original recommendation model: green indicates improvement, red indicates degradation, and unchanged results are left uncolored.
}
\centering
\label{tab:comp}
\small
\resizebox{\textwidth}{!}{%
\begin{tabular}{cr|ccc|cc|cr||ccc|cc|cr}
\toprule
\multicolumn{2}{c|}{Pretrained Model} & \multicolumn{7}{c||}{SASRec} & \multicolumn{7}{c}{FDSA} \\ \midrule
 & Metrics & Original & +FHFRS & +CFARS & \begin{tabular}[c]{@{}c@{}}+GRACE\\ {\footnotesize  (GradNorm)}\end{tabular} 
& \begin{tabular}[c]{@{}c@{}}+GRACE\\ {\footnotesize  (MGDA)}\end{tabular} & +GRACE & Rel. (\%) & Original & +FHFRS & +CFARS & \begin{tabular}[c]{@{}c@{}}+GRACE\\ {\footnotesize(GradNorm)}\end{tabular} & \begin{tabular}[c]{@{}c@{}}+GRACE\\ {\footnotesize  (MGDA)}\end{tabular} & +GRACE & Rel. (\%) \\ \midrule

\multirow{6}{*}{K=10} & Hit $\uparrow$
& 0.0332
& \cGFour{0.0338}
& \cROne{0.0310}
& \cGTwo{\underline{0.0374}}
& \cGThree{0.0373}
& \cGOne{\textbf{0.0378}}
& 13.86
& 0.0351
& 0.0351
& \cGFour{0.0354}
& \cGOne{\textbf{0.0405}}
& \cGTwo{\underline{0.0400}}
& \cGThree{0.0358}
& 1.99 \\

 & NDCG $\uparrow$
& 0.0184
& \cGFour{0.0185}
& \cROne{0.0176}
& \cGOne{\textbf{0.0205}}
& \cGTwo{\underline{0.0204}}
& \cGThree{0.0202}
& 9.78
& 0.0206
& 0.0206
& \cGThree{0.0207}
& \cGOne{\textbf{0.0224}}
& \cGTwo{\underline{0.0222}}
& \cROne{0.0205}
& -0.49 \\

 & EIS $\downarrow$
& 95.19
& \cROne{96.00}
& \cGFour{91.37}
& \cGTwo{\underline{81.70}}
& \cGOne{\textbf{81.52}}
& \cGThree{89.68}
& 5.79
& 90.93
& 90.93
& \cGTwo{\underline{88.97}}
& \cROne{94.55}
& \cRTwo{94.32}
& \cGOne{\textbf{78.96}}
& 13.17 \\

 & NIS $\uparrow$
& 34.59
& \cGFive{35.90}
& \cGFour{37.60}
& \cGThree{39.54}
& \cGTwo{\underline{39.83}}
& \cGOne{\textbf{43.40}}
& 25.45
& 34.04
& 34.04
& \cGFour{34.06}
& \cGTwo{\underline{37.62}}
& \cGThree{37.55}
& \cGOne{\textbf{43.19}}
& 26.86 \\

 & HMI$\uparrow$
& 42.49
& \cRTwo{42.29}
& \cROne{42.15}
& \cGTwo{\underline{44.78}}
& \cGOne{\textbf{44.83}}
& \cGThree{44.49}
& 4.71
& \underline{43.46}
& \underline{43.46}
& \underline{43.46}
& \cROne{42.86}
& \cRTwo{43.09}
& \cGOne{\textbf{48.53}}
& 11.67 \\

\midrule
\multirow{6}{*}{K=15} & Hit $\uparrow$
& 0.0419
& \cRTwo{0.0417}
& \cROne{0.0390}
& \cGThree{0.0484}
& \cGTwo{\underline{0.0485}}
& \cGOne{\textbf{0.0496}}
& 18.38
& 0.0459
& 0.0459
& \cROne{0.0452}
& \cGThree{0.0506}
& \cGTwo{\underline{0.0507}}
& \cGOne{\textbf{0.0513}}
& 11.76 \\

 & NDCG $\uparrow$
& 0.0207
& \cRTwo{0.0206}
& \cROne{0.0197}
& \cGOne{\textbf{0.0234}}
& \cGTwo{0.0233}
& \cGOne{\textbf{0.0234}}
& 13.04
& 0.0235
& 0.0235
& \cROne{0.0233}
& \cGOne{\textbf{0.0250}}
& \cGOne{\textbf{0.0250}}
& \cGTwo{0.0246}
& 4.68 \\

 & EIS $\downarrow$
& 104.16
& \cROne{104.48}
& \cGThree{89.32}
& \cGOne{\textbf{85.94}}
& \cGTwo{\underline{86.45}}
& \cGFour{101.82}
& 2.25
& 91.41
& \cGThree{91.32}
& \cGTwo{\underline{84.11}}
& \cROne{95.70}
& \cRTwo{95.23}
& \cGOne{\textbf{83.35}}
& 8.82 \\

 & NIS $\uparrow$
& \underline{38.68}
& \cRFour{38.51}
& \cROne{33.97}
& \cRTwo{37.95}
& \cRThree{38.18}
& \cGOne{\textbf{42.38}}
& 9.59
& 34.74
& \cRTwo{34.70}
& \cROne{33.01}
& \cGTwo{\underline{38.04}}
& \cGThree{37.97}
& \cGOne{\textbf{38.52}}
& 10.90 \\

 & HMI$\uparrow$
& 43.63
& \cRTwo{43.52}
& \cGFour{45.40}
& \cGTwo{\underline{45.86}}
& \cGThree{45.85}
& \cGOne{\textbf{47.08}}
& 7.91
& \underline{44.01}
& \cRFour{43.98}
& \cRThree{43.33}
& \cROne{42.73}
& \cRTwo{42.82}
& \cGOne{\textbf{47.91}}
& 8.87 \\

\midrule
\multirow{6}{*}{K=20} & Hit $\uparrow$
& 0.0531
& \cRTwo{0.0523}
& \cROne{0.0468}
& \cGThree{0.0554}
& \cGTwo{\underline{0.0557}}
& \cGOne{\textbf{0.0585}}
& 10.17
& 0.0553
& \cRTwo{0.0551}
& \cROne{0.0547}
& \cGThree{0.0580}
& \cGTwo{\underline{0.0582}}
& \cGOne{\textbf{0.0610}}
& 10.31 \\

 & NDCG $\uparrow$
& 0.0233
& \cRTwo{0.0231}
& \cROne{0.0216}
& \cGTwo{\underline{0.0251}}
& \cGThree{0.0250}
& \cGOne{\textbf{0.0255}}
& 9.44
& 0.0257
& \cRTwo{0.0256}
& \cROne{0.0255}
& \cGTwo{\underline{0.0268}}
& \cGTwo{\underline{0.0268}}
& \cGOne{\textbf{0.0269}}
& 4.67 \\

 & EIS $\downarrow$
& 95.89
& \cGFour{95.23}
& \cGThree{87.30}
& \cGOne{\textbf{83.13}}
& \cGTwo{\underline{83.51}}
& \cROne{99.02}
& -3.26
& 91.15
& \cGThree{90.64}
& \cGOne{\textbf{84.48}}
& \cROne{95.08}
& \cRTwo{94.94}
& \cGTwo{\underline{86.35}}
& 5.26 \\

 & NIS $\uparrow$
& \underline{38.98}
& \cRFour{38.84}
& \cROne{35.49}
& \cRTwo{35.84}
& \cGThree{35.94}
& \cGOne{\textbf{41.36}}
& 6.11
& 36.04
& \cRTwo{35.84}
& \cROne{34.14}
& \cGOne{\textbf{37.63}}
& \cGTwo{\underline{37.62}}
& \cGThree{37.51}
& 4.07 \\

 & HMI$\uparrow$
& 44.48
& \cRTwo{44.26}
& \cROne{43.36}
& \cGThree{45.68}
& \cGTwo{\underline{45.73}}
& \cGOne{\textbf{48.47}}
& 8.97
& \underline{43.76}
& \cRFour{43.58}
& \cROne{42.96}
& \cRThree{43.43}
& \cRTwo{43.42}
& \cGOne{\textbf{46.23}}
& 5.64 \\

\midrule\midrule
\multicolumn{2}{c|}{Pretrained Model}  & \multicolumn{7}{c||}{FEARec} & \multicolumn{7}{c}{GRAPE} \\ \midrule
 & Metrics & Original & +FHFRS & +CFARS & \begin{tabular}[c]{@{}c@{}}+GRACE\\ {\footnotesize  (GradNorm)}\end{tabular} 
& \begin{tabular}[c]{@{}c@{}}+GRACE\\ {\footnotesize  (MGDA)}\end{tabular} & +GRACE & Rel. (\%) & Original & +FHFRS & +CFARS & \begin{tabular}[c]{@{}c@{}}+GRACE\\ {\footnotesize(GradNorm)}\end{tabular} & \begin{tabular}[c]{@{}c@{}}+GRACE\\ {\footnotesize  (MGDA)}\end{tabular} & +GRACE & Rel. (\%) \\ \midrule

\multirow{6}{*}{K=10} & Hit $\uparrow$
& 0.0393
& \cRTwo{0.0392}
& \cROne{0.0383}
& \cGTwo{\underline{0.0397}}
& \cGOne{\textbf{0.0398}}
& 0.0393
& 0.00
& 0.0382
& \cROne{0.0381}
& \cGTwo{\underline{0.0390}}
& \cGThree{0.0384}
& \cGFour{0.0383}
& \cGOne{\textbf{0.0402}}
& 5.24 \\

 & NDCG $\uparrow$
& \textbf{0.0222}
& \textbf{0.0222}
& \cRTwo{0.0219}
& \cROne{0.0218}
& \cROne{0.0218}
& \cROne{0.0218}
& -1.80
& 0.0210
& 0.0210
& \cGTwo{\underline{0.0212}}
& \cROne{0.0208}
& \cROne{0.0208}
& \cGOne{\textbf{0.0217}}
& 3.33 \\

 & EIS $\downarrow$
& 106.71
& \cRTwo{106.73}
& \cROne{112.85}
& \cGThree{96.77}
& \cGTwo{\underline{96.68}}
& \cGOne{\textbf{96.52}}
& 9.55
& 112.63
& \cRThree{112.88}
& \cROne{118.65}
& \cGTwo{\underline{97.10}}
& \cGOne{\textbf{96.93}}
& \cRTwo{114.86}
& -1.97 \\

 & NIS $\uparrow$
& 40.48
& 40.48
& \cGTwo{\underline{40.71}}
& \cROne{40.08}
& \cROne{40.08}
& \cGOne{\textbf{42.56}}
& 5.14
& 40.60
& \cGFour{40.62}
& \cGThree{42.05}
& \cGTwo{\underline{42.31}}
& \cROne{37.77}
& \cGOne{\textbf{44.13}}
& 8.69 \\

 & HMI$\uparrow$
& 40.53
& \cGFive{40.55}
& \cGFour{41.71}
& \cGThree{44.55}
& \cGTwo{\underline{44.57}}
& \cGOne{\textbf{46.01}}
& 13.50
& 41.45
& \cGFive{41.47}
& \cGThree{43.33}
& \cGOne{\textbf{44.68}}
& \cGFour{42.12}
& \cGTwo{\underline{43.66}}
& 5.34 \\

\midrule
\multirow{6}{*}{K=15} & Hit $\uparrow$
& 0.0487
& 0.0487
& \cROne{0.0486}
& \cGOne{\textbf{0.0506}}
& \cGTwo{\underline{0.0505}}
& \cGThree{0.0494}
& 1.44
& 0.0484
& \cGTwo{\underline{0.0485}}
& \cROne{0.0475}
& \cRTwo{0.0476}
& \cRThree{0.0478}
& \cGOne{\textbf{0.0486}}
& 0.41 \\

 & NDCG $\uparrow$
& \textbf{0.0247}
& \textbf{0.0247}
& \cRTwo{0.0246}
& \cRTwo{0.0246}
& \cRTwo{0.0246}
& \cROne{0.0244}
& -1.21
& \underline{0.0237}
& \underline{0.0237}
& \cRTwo{0.0234}
& \cROne{0.0233}
& \cROne{0.0233}
& \cGOne{\textbf{0.0240}}
& 1.27 \\

 & EIS $\downarrow$
& 97.82
& \cROne{98.09}
& \cRTwo{106.52}
& \cGThree{96.54}
& \cGTwo{\underline{96.38}}
& \cGOne{\textbf{90.29}}
& 7.70
& 96.39
& \cRThree{99.24}
& \cROne{108.18}
& \cGOne{\textbf{85.70}}
& \cGTwo{\underline{94.06}}
& \cRTwo{101.12}
& -4.91 \\

 & NIS $\uparrow$
& 40.14
& \cGThree{40.20}
& \cGTwo{\underline{40.46}}
& \cRTwo{39.72}
& \cROne{39.68}
& \cGOne{\textbf{40.88}}
& 1.83
& 38.90
& \cGFour{39.52}
& \cGTwo{\underline{41.11}}
& \cGThree{39.66}
& \cROne{38.33}
& \cGOne{\textbf{42.30}}
& 8.76 \\

 & HMI$\uparrow$
& 41.99
& \cGFour{42.02}
& \cROne{41.09}
& \cGThree{44.22}
& \cGTwo{\underline{44.24}}
& \cGOne{\textbf{45.96}}
& 9.45
& 41.69
& \cGFive{41.89}
& \cGFour{42.23}
& \cGOne{\textbf{44.58}}
& \cGThree{42.48}
& \cGTwo{\underline{43.60}}
& 4.59 \\

\midrule
\multirow{6}{*}{K=20} & Hit $\uparrow$
& 0.0569
& \cGThree{0.0572}
& \cROne{0.0566}
& \cGTwo{\underline{0.0585}}
& \cGTwo{\underline{0.0585}}
& \cGOne{\textbf{0.0594}}
& 4.39
& 0.0560
& \cGOne{\textbf{0.0565}}
& \cRTwo{0.0555}
& \cGTwo{\underline{0.0563}}
& \cRThree{0.0557}
& \cROne{0.0551}
& -1.61 \\

 & NDCG $\uparrow$
& \underline{0.0267}
& \underline{0.0267}
& \cROne{0.0265}
& \cROne{0.0265}
& \cROne{0.0265}
& \cGOne{\textbf{0.0268}}
& 0.37
& \underline{0.0255}
& \cGOne{\textbf{0.0256}}
& \cRTwo{0.0253}
& \cRThree{0.0254}
& \cROne{0.0252}
& \underline{0.0255}
& 0.00 \\

 & EIS $\downarrow$
& 92.01
& \cRTwo{92.59}
& \cROne{101.12}
& \cGThree{90.36}
& \cGTwo{\underline{90.26}}
& \cGOne{\textbf{88.44}}
& 3.88
& \underline{86.56}
& \cRFour{89.12}
& \cROne{100.61}
& \cGOne{\textbf{82.27}}
& \cRThree{89.81}
& \cRTwo{92.55}
& -6.93 \\

 & NIS $\uparrow$
& 39.47
& \cGThree{39.59}
& \cGOne{\textbf{40.03}}
& \cRTwo{38.53}
& \cROne{38.52}
& \cGTwo{\underline{39.96}}
& 1.25
& 36.52
& \cGFour{37.13}
& \cGTwo{\underline{39.71}}
& \cGThree{37.98}
& \cGFour{37.13}
& \cGOne{\textbf{39.88}}
& 9.20 \\

 & HMI$\uparrow$
& 43.16
& \cGFour{43.22}
& \cROne{42.79}
& \cGTwo{\underline{45.34}}
& \cGOne{\textbf{45.35}}
& \cGThree{45.30}
& 4.97
& 42.67
& \cGFour{42.82}
& \cROne{42.48}
& \cGOne{\textbf{44.12}}
& \cGThree{43.39}
& \cGTwo{\underline{43.91}}
& 2.91 \\

\bottomrule
\end{tabular}
}
\end{table*}

\begin{table*}[t]
\caption{
Performance comparison on \textit{RecipeEmission}. “Rel.” denotes GRACE's relative improvement over the corresponding pretrained model. Best and second-best results are in bold and underlined, respectively. Cells are color-coded relative to the original pretrained model (green: improved, red: degraded, blank: unchanged).
}
\centering
\label{tab:re}
\small
\resizebox{\textwidth}{!}{%
\begin{tabular}{cr|ccc|cc|cc||ccc|cc|cc}
\toprule
\multicolumn{2}{c|}{Pretrained Model} 
& \multicolumn{7}{c||}{FEARec} 
& \multicolumn{7}{c}{FDSA} \\ 
\midrule
& Metrics 
& Original 
& +FHFRS 
& +CFARS 
& \begin{tabular}[c]{@{}c@{}}+GRACE\\ {\footnotesize(GradNorm)}\end{tabular}
& \begin{tabular}[c]{@{}c@{}}+GRACE\\ {\footnotesize(MGDA)}\end{tabular}
& +GRACE 
& Rel. (\%) 
& Original 
& +FHFRS 
& +CFARS 
& \begin{tabular}[c]{@{}c@{}}+GRACE\\ {\footnotesize(GradNorm)}\end{tabular}
& \begin{tabular}[c]{@{}c@{}}+GRACE\\ {\footnotesize(MGDA)}\end{tabular}
& +GRACE 
& Rel. (\%) \\ 
\midrule

\multirow{3}{*}{K=10} 
& Hit $\uparrow$
& 0.0391
& 0.0391
& \cROne{0.0390}
& \cGOne{\textbf{0.0397}}
& \cGTwo{\underline{0.0396}}
& \cROne{0.0390}
& -0.26
& \textbf{0.0428}
& \textbf{0.0428}
& \textbf{0.0428}
& \cROne{\underline{0.0427}}
& \cRTwo{0.0425}
& \cRThree{0.0417}
& -2.57 \\

& NDCG $\uparrow$
& \textbf{0.0192}
& \cROne{0.0191}
& \textbf{0.0192}
& \textbf{0.0192}
& \textbf{0.0192}
& \textbf{0.0192}
& 0.00
& \underline{0.0205}
& \underline{0.0205}
& \underline{0.0205}
& \underline{0.0205}
& \cROne{0.0204}
& \cGOne{\textbf{0.0208}}
& 1.46 \\

& CO$_2$ $\downarrow$
& 12.84
& \cGThree{12.57}
& \cGFour{12.66}
& \cGFive{12.82}
& \cGTwo{\underline{12.51}}
& \cGOne{\textbf{12.12}}
& 5.66
& 15.50
& \cGFive{15.44}
& \cGFive{15.44}
& \cGTwo{\underline{14.69}}
& \cGThree{15.04}
& \cGOne{\textbf{13.36}}
& 13.83 \\

\midrule

\multirow{3}{*}{K=15} 
& Hit $\uparrow$
& \textbf{0.0561}
& \cRTwo{0.0558}
& \cROne{\underline{0.0560}}
& \cRThree{0.0557}
& \cRThree{0.0557}
& \cRTwo{0.0558}
& -0.53
& 0.0553
& 0.0553
& 0.0553
& \cGOne{\textbf{0.0566}}
& \cGThree{0.0555}
& \cGTwo{\underline{0.0557}}
& 0.72 \\

& NDCG $\uparrow$
& \textbf{0.0237}
& \cRTwo{0.0235}
& \textbf{0.0237}
& \cRThree{0.0234}
& \cRThree{0.0234}
& \textbf{0.0237}
& 0.00
& 0.0238
& 0.0238
& 0.0238
& \cGTwo{\underline{0.0242}}
& \cGThree{0.0239}
& \cGOne{\textbf{0.0245}}
& 2.94 \\

& CO$_2$ $\downarrow$
& 14.60
& \cGFour{14.30}
& \cGFive{14.57}
& \cGTwo{\underline{14.01}}
& \cGOne{\textbf{13.91}}
& \cGThree{14.27}
& 2.27
& 15.16
& \cGFive{14.96}
& \cGFive{14.96}
& \cGTwo{\underline{13.03}}
& \cGOne{\textbf{12.68}}
& \cGThree{13.78}
& 9.12 \\

\midrule

\multirow{3}{*}{K=20} 
& Hit $\uparrow$
& 0.0703
& \cGOne{\textbf{0.0704}}
& \cGOne{\textbf{0.0704}}
& \cRTwo{0.0692}
& \cROne{0.0691}
& \cGOne{\textbf{0.0704}}
& 0.14
& 0.0661
& 0.0661
& 0.0661
& \cGOne{\textbf{0.0671}}
& \cGTwo{\underline{0.0669}}
& \cGThree{0.0665}
& 0.61 \\

& NDCG $\uparrow$
& \textbf{0.0271}
& \cRTwo{0.0269}
& \textbf{0.0271}
& \cRThree{0.0266}
& \cRThree{0.0266}
& \textbf{0.0271}
& 0.00
& 0.0264
& 0.0264
& 0.0264
& \cGTwo{\underline{0.0266}}
& \cGThree{0.0265}
& \cGOne{\textbf{0.0271}}
& 2.65 \\

& CO$_2$ $\downarrow$
& 13.25
& \cGFour{13.23}
& 13.25
& \cGTwo{\underline{12.92}}
& \cGOne{\textbf{12.90}}
& \cGThree{13.19}
& 0.47
& 14.34
& \cGFive{14.26}
& \cGFive{14.26}
& \cGTwo{\underline{12.73}}
& \cGThree{13.04}
& \cGOne{\textbf{12.18}}
& 15.06 \\

\bottomrule
\end{tabular}
}
\end{table*}

% \begin{figure*}[t]
%     \centering
%     \subfigure[Projection Ratio]{
%         \includegraphics[width=0.3\textwidth]{figures/proj.pdf}
%     }
%     \hfill
%     \subfigure[Top-K for Differentiable Loss]{
%         \includegraphics[width=0.3\textwidth]{figures/topk.pdf}
%     }
%     \hfill
%     \subfigure[Temperature in Gumbel-Softmax]{
%         \includegraphics[width=0.3\textwidth]{figures/tau.pdf}
%     }
%     \caption{Hyperparameter Analysis in GRACE}
%     \label{fig:hyperparam_analysis}
% \end{figure*}

\section{Experiments}
\label{sec:expert}
\subsection{Experimental Settings}
\subsubsection{Pretrained Recommendation Models} 
% To systematically evaluate the effectiveness of GRACE in enhancing both accuracy and sustainability, we select four representative sequential recommendation models as base architectures and fine-tune them using GRACE.
% To ensure that GRACE is evaluated on a broad and representative set of recommendation models, we select four architectures that reflect the main ways user preference modelling is carried out in current research and practice. SASRec serves as a simple and widely adopted attention-based model, making it a natural starting point for examining how GRACE behaves with standard scoring mechanisms. FDSA introduces item features into the modeling process, allowing us to test GRACE in settings where user choices are influenced not only by past interactions but also by semantic attributes of items. FEARec captures behavioral signals that appear with certain rhythms or frequencies, providing a different perspective on how users form preferences over time. In addition to these general-purpose models, we include GRAPE, which already incorporates sustainability-related objectives. Using GRACE on top of GRAPE allows us to examine whether our approach can further strengthen green performance even when the baseline already considers environmental factors.

To evaluate GRACE on a broad and representative set of recommendation models, we fine-tune it on four widely used or state-of-the-art neural backbones that learn user preferences from historical interactions. Our goal is to assess whether GRACE can improve sustainability while preserving accuracy across diverse architectures and scoring mechanisms. Specifically, we include an attention-based baseline, a feature-aware backbone, a model with an alternative signal view, and a sustainability-aware recommender, covering complementary inductive biases and design choices.

\begin{itemize}
\item \textbf{SASRec}~\cite{kang2018self} is a widely adopted self-attention-based recommender that models dependencies in a user’s interaction history using a uni-directional attention encoder with positional encoding and masking.

% models user behavior sequences with a uni-directional self-attention encoder.
% It learns the dependency between a target item and all previous interactions by assigning attention weights, and it uses position embeddings and masking to preserve temporal order. 

\item \textbf{FDSA}~\cite{zhang2019feature} extends attention-based recommendation by incorporating item features via two coordinated attention modules that capture item transitions and feature representations, respectively.

% extends attention-based recommendation by introducing two coordinated attention modules.
% One module captures transition patterns among items, while the other attends to item feature representations. These two attention outputs are then combined, enabling the model to use both behavior dynamics and descriptive item information when producing predictions.

\item \textbf{FEARec}~\cite{du2023frequency} introduces a frequency-aware attention mechanism based on the discrete Fourier transform, modeling interaction signals in the frequency domain to capture periodic preference patterns.

% incorporates a frequency-aware attention layer built on top of the discrete Fourier transform. It decomposes user behavior signals into different frequency components and applies attention in the frequency domain. 

\item \textbf{GRAPE}~\cite{jing2025bites} explicitly models sustainability by integrating nutrition and environmental indicators into its scoring and training objectives, balancing preference with greener choices.

% is constructed with a dedicated sustainability modeling pathway in addition to its standard preference modeling components. It integrates nutrition and environmental indicators into its scoring function and applies a training objective that balances user preference with greener choices. GRAPE thus produces rankings that reflect both personal taste and sustainability considerations.
\end{itemize}

Together, these backbones form a strong testbed for benchmarking GRACE as a general sustainability-aware fine-tuning framework.

\subsubsection{Dataset and Evaluation Metrics}
We evaluate GRACE on two real-world datasets, \textit{GreenRec}~\footnote{\url{drive.google.com/drive/folders/11cdceu3Z2e-3NKzEVI6qoU63lZFxTo8Y?usp=sharing}} and \textit{RecipeEmission}~\footnote{\url{github.com/RaoulKalisvaart/green-recommender-systems}}, using metrics that capture both recommendation accuracy and sustainability or health outcomes. The distributions of the sustainability indicators are shown in Figure~\ref{fig:data}.

% To evaluate the effectiveness of GRACE, we conduct experiments on two real-world datasets: \textit{GreenRec}~\footnote{drive.google.com/drive/folders/11cdceu3Z2e-3NKzEVI6qoU63lZFxTo8Y?usp=sharing} and \textit{RecipeEmission}~\footnote{https://github.com/RaoulKalisvaart/green-recommender-systems}, using metrics that capture both recommendation accuracy and sustainability or health-related performance. The distributions of sustainability indicators across the two datasets are illustrated in Figure~\ref{fig:data}. 

\textit{GreenRec} extends public recipe data with rich user interaction histories and item-level sustainability annotations. Each recipe is associated with three indicators: Environmental Impact Score (EIS), and two health-oriented indicators, Nutritional Impact Score (NIS) and Healthy Meal Index (HMI)~\cite{zhang2024greenrec}. We report standard recommendation accuracy metrics, Hit Ratio@K and NDCG@K. To evaluate sustainability and health alignment, we compute the average indicator values among the top-K recommended items, namely EIS@K, NIS@K, and HMI@K. Higher NIS and HMI indicate healthier recommendations, while EIS reflects environmental burden and is better when lower.

\textit{RecipeEmission} focuses on environmental sustainability by estimating each recipe's carbon footprint using ingredient-level emission factors~\cite{kalisvaart2025towards}. The CO$_2$ distribution is long-tailed as shown in Figure~\ref{fig:data}(b). We evaluate accuracy using Hit Rate@K and NDCG@K, and report the average CO$_2$-eq (CO$_2$) among the top-$K$ recommended items, where lower values indicate better environmental sustainability. We use \textit{RecipeEmission} as a CO$_2$-oriented validation dataset. Since the main \textit{GreenRec} experiments already cover four backbones, we use FEARec and FDSA as representative sequential backbones for the \textit{RecipeEmission} evaluation.

\begin{figure}[t]
    \centering
    \includegraphics[width=.96\linewidth]{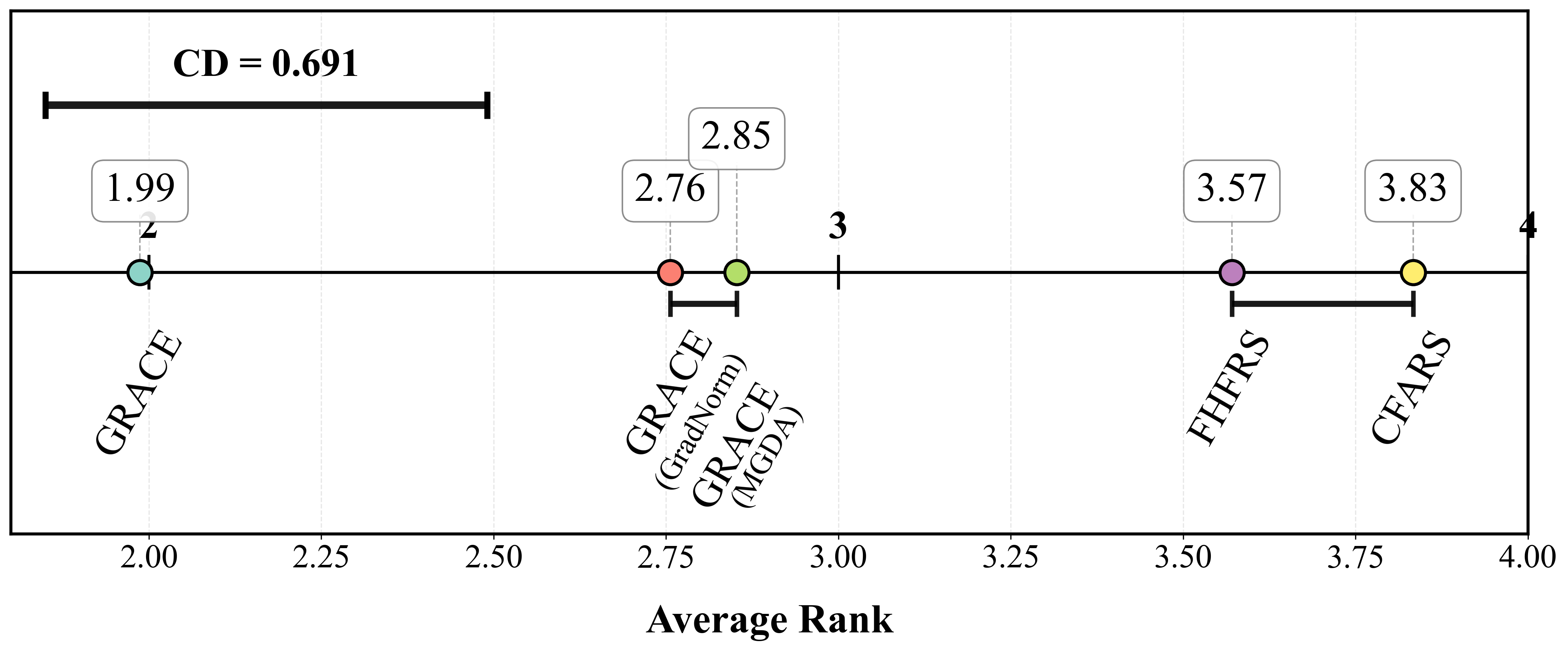}
    \caption{Critical Difference (CD) diagram based on average ranks across datasets using the Friedman test with Nemenyi post-hoc ($p<0.05$). The x-axis shows average rank (smaller is better). Horizontal bars connect methods with no significant pairwise differences; within each group, the leftmost method ranks best.
    % Critical Difference (CD) Diagram. Average ranks across datasets are compared using the Friedman test with Nemenyi post-hoc; differences are significant when indicated by p-values from the Nemenyi post-hoc at the 0.05 significance level. The x-axis shows average ranks (smaller is better; 1 at the left). Horizontal bars indicate groups with non-significant pairwise differences; within a group, the leftmost method is best ranked.
    }
    \label{fig:cd}
\end{figure}

\begin{figure*}[t]
    \centering
    \includegraphics[width=0.98\textwidth]{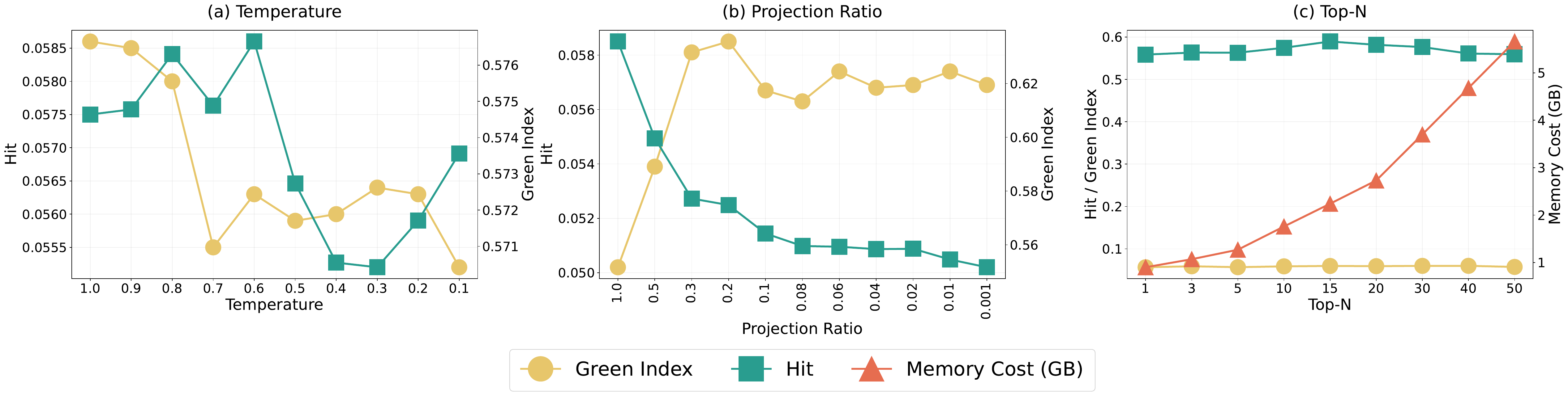}
    \caption{Hyperparameter Analysis in GRACE}
    \label{fig:hyperparam_analysis}
\end{figure*}

\subsubsection{Baseline Methods}
To benchmark GRACE, we compare against strong reranking-based methods that incorporate item-level sustainability signals at inference time, as well as standard multi-objective optimization strategies.

% To benchmark our proposed method, we select the below strong reranking-based baselines. Both baselines could explicitly incorporate sustainability or health-awareness into recommendation.

\begin{itemize}
\item \textbf{FHFRS}~\cite{rostami2023towards} is a penalty-based reranking method originally proposed for exposure fairness. We adapt it to sustainability by partitioning items into \emph{green} and \emph{non-green} groups using a binary threshold on an item-level green metric (e.g., median split on an environmental score). It then penalizes items in the low-green group during reranking:
% adopts a penalty-based reranking strategy to ensure exposure fairness across green and non-green items. It assigns binary labels to items based on a predefined green metric (e.g., EIS above median) and penalizes items with low green scores by subtracting a fixed value. The final reranked scores $s_i^{new}$ for each item $i$ are computed as:
\begin{equation}
s_i^{new} = \gamma \cdot s_i - \lambda \cdot \mathbb{I}(\text{low-green}_i),
\end{equation}
where $s_i$ is the base model score, $\mathbb{I}(\cdot)$ indicates whether item $i$ falls below the green threshold, $\lambda$ controls the penalty strength, and $\gamma$ rescales the original scores.
% This approach enhances the exposure of green items while maintaining recommendation quality without model retraining.

\item \textbf{CFARS}~\cite{kalisvaart2025towards} performs score-level fusion between the base model relevance score and an item-level sustainability score:
% linearly fuses sustainability scores with model prediction scores:
\begin{equation}
s_i^{new} = \delta \cdot s_i + (1 - \delta) \cdot g_i,
\end{equation}
where $g_i$ is a normalized green score (e.g., CO$_2$-derived greenness in \textit{RecipeEmission}, or a weighted combination of EIS, NIS, and HMI in \textit{GreenRec}). The trade-off parameter $\delta \in [0,1]$ controls the emphasis on preference versus sustainability.
% This method encourages high-scoring green items to rank higher in the recommendation list without retraining the base modelation.
\end{itemize}
% We select these two baselines as they represent typical reranking-based approaches that explicitly incorporate sustainability or health-awareness into recommendation. Both operate solely at evaluation time without requiring model retraining, making them efficient and practical for real-world deployment. Their lightweight design and clearly defined objectives provide a strong reference point for comparison against our fine-tuning-based framework.

Beyond inference-time reranking, we further compare a set of multi-objective optimization schemes to balance preference and sustainability objectives during model fine-tuning.
Specifically, we replace GRACE’s gradient projection module with \textbf{GradNorm}~\cite{chen2018gradnorm} and \textbf{Pareto MGDA}~\cite{sener2018multi}.
GradNorm adaptively reweights losses by matching gradient magnitudes, while Pareto MGDA computes a convex combination of task gradients that follows a Pareto-balanced descent direction. We denote these variants as \textbf{GRACE(GradNorm)} and \textbf{GRACE(MGDA)}. 
The reranking methods, \textbf{FHFRS} and \textbf{CFARS}, serve as direct green recommendation baselines. They are applied to the same candidate set produced by the pretrained backbone, and we report both accuracy and sustainability metrics after reranking. For FHFRS, the green/non-green split is computed per dataset using the corresponding item-level sustainability signal.

\begin{table}[t]
\caption{Ablation study of GRACE under different loss configurations. Cells shaded in greener tones indicate better performance, while redder tones denote worse outcomes. Cell colors indicate relative performance within each metric column (greener denotes better results and redder denotes worse ones).}
\label{tab:abla}
\centering
\resizebox{0.9\linewidth}{!}{
\begin{tabular}{lccccc}
\toprule
\textbf{Loss} 
& \textbf{Hit $\uparrow$} 
& \textbf{NDCG $\uparrow$} 
& \textbf{EIS $\downarrow$} 
& \textbf{NIS $\uparrow$} 
& \textbf{HMI$\uparrow$} \\
\midrule
\textbf{R} 
& \cGood{0.0572}
& \cBest{0.0260}
& \cBad{95.39}
& \cBad{37.77}
& \cBad{46.65} \\
\textbf{G} 
& \cBad{0.0027}
& \cBad{0.0008}
& \cBest{58.29}
& \cBest{50.89}
& \cBest{68.34} \\
\textbf{R + G'} 
& \cWorst{0.0018}
& \cWorst{0.0006}
& \cGood{58.85}
& \cGood{50.39}
& \cGood{68.00} \\
\textbf{R + G''} 
& \cGood{0.0572}
& \cBest{0.0260}
& \cBad{95.39}
& \cBad{37.77}
& \cBad{46.65} \\
\textbf{R' + G} 
& \cMid{0.0557}
& \cMid{0.0250}
& \cMid{83.88}
& \cWorst{35.94}
& \cWorst{45.71} \\
\textbf{R + G} 
& \cBest{0.0585}
& \cGood{0.0255}
& \cWorst{99.02}
& \cMid{41.36}
& \cMid{48.47} \\
\bottomrule
\end{tabular}
}
\end{table}

\subsubsection{Implementations}
All components are trained on a single NVIDIA RTX 3090 GPU with 24 GB RAM. All pretrained recommendation models are implemented using the RecBole framework. For data preprocessing, we retain users with interaction counts in $[5,50]$ and items with at least $2$ interactions. Each user's interaction sequence is chronologically split into training, validation, and test sets with proportions of $50\%$, $20\%$, and $30\%$, respectively. All experiments use a fixed random seed of 2025.

For the sustainability score used in optimization, we align all green indicators so that larger values indicate greener items. Each raw indicator $x_{i,m}$ is first min-max normalized as
\begin{equation}
\bar{x}_{i,m}=\frac{x_{i,m}-\min(x_m)}{\max(x_m)-\min(x_m)+\epsilon}.
\end{equation}
For lower-is-better indicators such as EIS and CO$_2$, we use $1-\bar{x}_{i,m}$; for higher-is-better indicators such as NIS and HMI, we directly use $\bar{x}_{i,m}$. On \textit{GreenRec}, the scalar score used in Eq.~(\ref{eq:lgreen}) is computed by averaging the aligned EIS, NIS, and HMI scores. On \textit{RecipeEmission}, we use the aligned CO$_2$-derived greenness score. The original indicators are still reported separately for evaluation.

For GRACE, we set the number of top-ranked positions in the differentiable loss to $N=10$ and the temperature to $\tau=0.9$. Model selection and hyperparameter tuning are based on the validation set using NDCG@20. On \textit{GreenRec}, FHFRS uses $\lambda=0.1$ and $\gamma=0.7$, while CFARS uses $\delta=0.8$. On \textit{RecipeEmission}, FHFRS uses $\lambda=0.05$ and $\gamma=0.95$, while CFARS uses $\delta=0.95$. GRACE is implemented on top of the original training pipelines of the pretrained models.

% Beyond reranking-based methods, we further compare against two strong multi-objective optimization baselines, \textbf{GradNorm}~\cite{chen2018gradnorm} and \textbf{Pareto MGDA}~\cite{sener2018multi}.
% GradNorm adaptively re-weights task losses by matching their gradient magnitudes, so that all objectives train at similar rates and no single loss dominates optimization.
% Pareto MGDA views multi-task learning as multi-objective optimization and, at each step, finds a convex combination of task gradients that yields a balanced descent direction aligned with the Pareto front.
% To ensure fairness, we integrate these optimization schemes into the GRACE framework by replacing its gradient-projection module. We denote the resulting variants as \textbf{GRACE(GradNorm)} and \textbf{GRACE(MGDA)}. 

\begin{figure}[t]
    \centering
    \includegraphics[width=\linewidth]{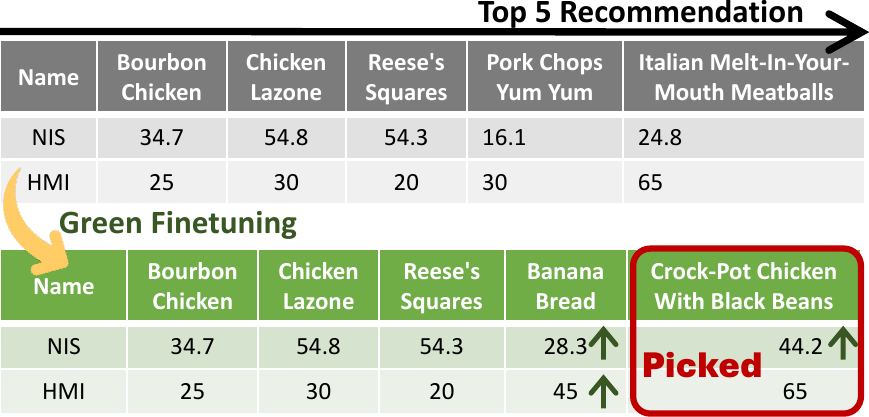}
    \caption{Case study of GRACE on \textit{GreenRec}. Top: Top-5 recipes from pretrained SASRec. Bottom: Top-5 recipes after GRACE fine-tuning. ``Picked'' indicates that the user interacted with the corresponding item.
    % Case analysis for GRACE. The left part shows the original recommended recipes by SASRec, and the right part lists the final recommendations after green finetuning. The centre part presents NIS and HMI for the top-5 recipes and their overall average, visualized as 100\% stacked horizontal bars that compare the original scores (grey) with the scores after green finetuning (green) for each rank position and for the average row.
    }
    \label{fig:case}
\end{figure}

\subsection{Results and Analysis}
\subsubsection{Overall Performance}
Table~\ref{tab:comp} and Table~\ref{tab:re} compare GRACE with the pretrained backbones and strong reranking baselines across the evaluated recommendation models. The primary goal is to improve sustainability while avoiding accuracy loss. Across most settings, GRACE achieves this goal by improving sustainability with comparable accuracy, and in several cases it improves both sustainability and accuracy, suggesting that sustainability-aware fine-tuning can also act as a helpful regularizer rather than a pure trade-off. 
For example, on \textit{GreenRec} with SASRec, GRACE improves Hit@15 and NDCG@15 by 18.38\% and 13.04\%, respectively, while also improving the sustainability scores.

In a few cases, accuracy decreases slightly, such as a small drop in NDCG@10 for FEARec on \textit{GreenRec}. Since GRACE optimizes a scalar sustainability score during training, individual sustainability indicators may also exhibit local trade-offs when environmental and health-related dimensions are not perfectly aligned. On \textit{RecipeEmission}, GRACE reduces the average CO$_2$ of recommended items on both FEARec and FDSA with limited impact on recommendation accuracy.

Although GRACE is a fine-tuning method and therefore does not aim for dramatic changes, the improvements are consistent and statistically reliable. The Critical Difference (CD) diagram in Figure~\ref{fig:cd} summarizes average ranks and statistical significance under the Nemenyi test. GRACE attains the best overall rank (1.99) and significantly outperforms all compared baselines, highlighting its robustness.

\subsubsection{Hyperparameter Analysis}
To evaluate GRACE's sensitivity, we analyze three key hyperparameters: the Gumbel-softmax temperature $\tau$ (Eq. (\ref{eq:topk})), the green loss projection ratio (Eq. (\ref{eq:finalgrad})), and the top-N items used for differentiable green loss.
To evaluate overall sustainability, we use a Green Index (GRI) as the average of the normalized values of all three indicators. Scores are scaled to the [0,1] range, and EIS is negated before normalization so that higher values consistently indicate better sustainability. We report Hit@20 for accuracy and GRI@20 for sustainability using SASRec as the pretrained recommendation model on the \textit{GreenRec} dataset.

Figure~\ref{fig:hyperparam_analysis}(a) illustrates how GRACE responds to changes in the temperature parameter $\tau$. Higher values of $\tau$ produce smoother approximations, which improve recommendation accuracy (Hit@20). When $\tau = 0.8$ and $\tau = 0.9$, both accuracy and sustainability remain relatively high.  In contrast, the sustainability metric (GRI@20) peaks around $\tau=0.6$ and then fluctuates as the temperature decreases. These results highlight the importance of selecting an appropriate temperature value to strike a balance between accuracy and sustainability.

As shown in Figure~\ref{fig:hyperparam_analysis}(b), the projection ratio controls the trade-off between sustainability and recommendation accuracy. A higher projection ratio significantly boosts the recommendation sustainability but may reduce accuracy due to overly aggressive green-oriented updates. In contrast, a lower projection ratio helps maintain accuracy but leads to weaker sustainability performance. This suggests that the projection ratio can serve as a controllable knob for green fine-tuning, where smaller values provide more conservative preference preservation and larger values allow stronger sustainability-oriented adaptation.

Figure~\ref{fig:hyperparam_analysis}(c) shows how the choice of top-N items for computing the differentiable green loss affects performance. Both GRI@20 and Hit@20 reach their best values around N = 15 and remain stable across moderate N values. However, GPU memory usage increases rapidly with larger N, growing from 0.9 GB when N = 1 to 5.66 GB when N = 50.

\subsubsection{Ablation Study}
To understand the contribution of each component in GRACE, we perform an ablation study using SASRec as the pretrained model on the \textit{GreenRec} dataset. Table~\ref{tab:abla} reports the results of several key variants: R refers to GRACE’s differentiable recommendation loss; R' uses the original loss from the pretrained model; G is the differentiable green loss with gradient projection; G' is the same green loss but without projection; and G'' adds a non-differentiable green score directly to the loss.

% To better understand the contributions of each component within GRACE, we conduct an ablation study on SASRec at K=20. Table~\ref{tab:abla} summarizes the performance of several key variants: R denotes the differentiable recommendation loss within GRACE; R' represents the original base model loss; G indicates the differentiable green loss with gradient projection onto the recommendation gradient direction; G' denotes the differentiable green loss without gradient projection; and G'' represents a non-differentiable green score directly added to the loss.

We make the following observations. First, the full GRACE model (R + G) achieves the most balanced performance, demonstrating the effectiveness of training with both differentiable recommendation loss and projected green loss. Compared to using the projected green loss alone (G), adding the recommendation objective (R + G) significantly improves accuracy, with Hit@20 increasing from 0.0027 to 0.0585. Second, removing the gradient projection (R + G') yields similarly high green scores but severely degrades accuracy, as Hit@20 drops to 0.0018. This highlights the importance of projection in preserving recommendation quality. Third, incorporating non-differentiable green scores (R + G'') fails to improve over the baseline, indicating that differentiability is essential for effective optimization. Moreover, replacing GRACE’s recommendation loss with the pretrained recommendation model loss (R' + G) results in comparable green performance but remains inferior to the full GRACE model. Overall, these results demonstrate that GRACE’s design, which combines differentiable objectives with gradient alignment, is key to achieving sustainability gains without compromising accuracy.

% Results demonstrate that the full GRACE design (R + G) achieves the most balanced performance, validating the efficient functionality of each component. For example, compared to applying projected green loss alone (G), the combined (R + G) significantly improves recommendation accuracy (Hit@20 from 0.0027 to 0.0585), although it sacrifices a portion of green performance (GRI drops from 0.7453 to 0.5748). Besides, directly combining differentiable green loss without projection (R + G') achieves similarly high GRI (0.7410) but severely reduces Hit@20 (0.0018), underscoring the importance of projection to maintain recommendation fidelity. Incorporating non-differentiable green scores (R + G'') yields no improvement over baseline (Hit@20 = 0.0572, GRI = 0.5562), indicating that differentiability is essential for effective green signal integration. Moreover, using the original recommendation loss with projected green loss (R' + G) provides moderate improvements (GRI = 0.5593) but remains inferior to GRACE’s fully integrated formulation. These results confirm that GRACE’s design, differentiable formulation plus gradient alignment, is key to achieving robust sustainability gains without compromising recommendation quality.

\subsubsection{Case Study}
As shown in Figure~\ref{fig:case}, this case study illustrates how GRACE can improve recommendation outcomes in a preference-preserving way. After fine-tuning, the recommended list shifts toward recipes with higher health-related scores (NIS and HMI), while the top-ranked items remain largely stable. Notably, in this example the pretrained SASRec recommendations contain no picked items, whereas the GRACE-tuned list includes one picked item, suggesting that incorporating health-aware objectives can improve both health alignment and relevance for some users.

\section{Conclusion}
\label{sec:conclusion}

In conclusion, this paper presents GRACE, an efficient fine-tuning framework for integrating sustainability objectives into existing recommender systems without the need for full retraining. GRACE jointly optimizes for accuracy and sustainability by introducing a differentiable green loss combined with a gradient projection mechanism. This design overcomes the challenge of non-differentiable green objectives and mitigates conflicts between sustainability supervision and user preference learning. Future work will explore more fine-grained user modeling that captures individual attitudes toward sustainability, enabling personalized calibration of green goals. 
% Additionally, we plan to extend our evaluation to more diverse datasets that include both user-item interactions and item-level green attributes, further validating GRACE’s applicability across domains. 
Still, we believe this work offers a practical and efficient approach to green recommendation, with the potential to guide consumers toward more sustainable and healthier lifestyles.

% \section*{Acknowledgment}
\section*{Acknowledgments}
This research is supported in part by the RIE2025 Industry Alignment Fund -- Industry Collaboration Projects (IAF-ICP) (Award I2301E0026), administered by A*STAR; Alibaba Group and NTU Singapore through the Alibaba-NTU Global e-Sustainability CorpLab (ANGEL); the Joint NTU-UBC Research Centre of Excellence in Active Living for the Elderly (LILY), Nanyang Technological University, Singapore; the National Natural Science Foundation of China (No. 92367202, No. 62202279); the Excellent Youth Science Fund Project (Overseas) of Shandong (No. 2023HWYQ-039); the Natural Science Foundation of Shandong Province (No. ZR2022QF018, No. ZR2023LZH006); and the Fundamental Research Funds of Shandong University.

\ifCLASSOPTIONcaptionsoff
  \newpage
\fi

\newpage

\normalem
\bibliographystyle{IEEEtran}
\bibliography{TKDE}

\end{document}